# Half-Wave Dipolar Metal-Semiconductor Laser

Sangyeon Cho[1,2], Nicola Martino[1,2], Seok-Hyun Yun[1-3]*

[1]Wellman Center for Photomedicine, Massachusetts General Hospital, 65 Landsdowne St., Cambridge, MA 02139, USA
[2]Harvard Medical School, Boston, MA 02139, USA
[3]Harvard-MIT Health Sciences and Technology, Cambridge, MA 02139, USA
*Correspondence to: syun@hms.harvard.edu

**Abstract**

Nano-scale lasers harnessing metallic plasmons hold promise across physical sciences and industrial applications. Plasmons are categorized as surface plasmon polaritons (SPP) and localized surface plasmons (LSP).[1] While SPP has gained popularity for nano-lasers[2,3] by fitting a few cycles of SPP waves into resonators, achieving LSP lasing in single nanoparticles remains an elusive goal.[4–6] Here, we highlight the equivalence of LSP and SPP within resonant systems and present lasers oscillating in the lowest-order LSP or, equivalently, half-cycle SPP. This diffraction-limited dipolar emitter is realized through strong coupling of plasmonic oscillation in gold and dielectric resonance in high-gain III-V semiconductor in the near infrared away from surface plasmon frequencies. The resulting single-mode stimulated emission peak exhibits linewidth Q factors over 50 at room temperature, with wide tunability spanning from 1190 to 1460 nm determined by resonator sizes ranging from 190 to 280 nm. A semiconductor laser model elucidates the temporal and spectral buildup dynamics under optical pumping. Notably, linewidth Q values surpassing 250 are attained from higher-order, isolated laser particles within live biological cells. These results offer fresh perspectives in nanophotonics and indicate promising opportunities for multiplexed biological applications.[2,7–9]

**Main text**

Every resonant structure, such as acoustic instruments, possess a fundamental oscillation frequency, whose half or quarter cycle precisely fits into the resonator. In optical lasers, this corresponds to a mode with a wavelength ($\lambda$) equal to half or a quarter of the resonator's length divided by the refractive index ($n$). At a given wavelength, the smallest laser size could be achieved by employing the lowest-order mode. To the best of our knowledge, however, laser oscillation in the sub-wavelength modes in all three dimensions has yet to be demonstrated.

Reaching the laser threshold, where optical gain equals loss, necessitates a pumping rate exceeding $M\omega/Q$, where $M$ denotes the number of modes to which pump energy is distributed (spontaneous emission factor $\beta = M^{-1}$), $\omega$ is the optical frequency, and $Q$ is the cavity quality factor. Furthermore, a sufficient number of gain elements is required to absorb the pump energy, at least $\omega\tau_s/Q$ emitters (see Supplementary Note 1), where $\tau_s^{-1}$ is the spontaneous radiative rate per mode. $\tau_s$ is modified by the Purcell effect from its free-space value $\tau_0$ by a Purcell factor $F_p = \frac{3\tau_0}{4\pi^2}\frac{Q}{V_m}\left(\frac{\lambda}{n}\right)^3$ with $V_m$ denoting the mode volume.[10] When the gain bandwidth (Δω) is broader than the cavity bandwidth ($\omega/Q$), the minimal emitter density, $\rho_{min}$, must exceed $\sim \frac{\Delta\omega\tau_0}{Q}\left(\frac{\lambda}{2n}\right)^{-3}$ (see Supplementary Note 1). For instance, for single-mode resonators with Δω= $3 \times 10^{14}$ s⁻¹, $\tau_0 = 10^{-9}$ s, $\lambda/2n$= 180 nm, and $Q = 10$, $\rho_{min} \approx 5 \times 10^{18}$ cm⁻³ is derived. Achieving this density with organic fluorophores would require a high concentration of 8.5 mM without nonradiative quenching. Moreover, deep-subwavelength metallic nanospheres demand excessive pump energy due to their large $M$ (> 40), which heats up the gain medium and metal. A previous observation of spectral narrowing from dye-doped nanospheres is attributed to a collective action of many nanoparticles.[5,6,11] Besides dyes, this minimum density is difficult to achieve with quantum dot emitters as $5 \times 10^{18}$ cm⁻³ translates to tight packing at a 5.2 nm spacing without Auger recombination loss.[12] On the other hand, successful nano-lasers have harnessed high-gain semiconductors and employed resonator lengths longer than the intracavity half wavelength ($\lambda/2n$), making use of higher order modes with relatively high $Q$ (>100). For example, photonic crystal lasers generate small mode volume at the cost of multi-wavelength resonator sizes to achieve high-Q Bragg reflection.[13] A coaxial metallic nano-laser has used a mode order of 2, the lowest mode order ever demonstrated.[14]

Our strategy to achieve fundamental-mode oscillation is as follows. Firstly, we target the half-wavelength mode in a fundamental resonator of a length $\frac{\lambda}{2n}$. A plasmonic cavity is essential since dielectric-only cavities offer near unity $Q$ as the cavity volume approaches or even gets smaller than $\left(\frac{\lambda}{2n}\right)^3$. Secondly, it is critical to minimize non-radiative coupling of emitters to higher order plasmonic modes. We show that $M = 1$ is attainable in the near infrared (NIR) away from surface plasmon resonance frequencies ($\omega_{SP}$). Thirdly, we employ III-V semiconductor driven at high-density (gain) states substantially above the band gap. The outcome is a single lasing mode characterized as a half-wave plasmonic dipole.

Figure 1a illustrates the plasmonic modes of gold nanospheres immersed in a dielectric medium with a refractive index ($n_m$) of 2.5 (representing the effects of III-V gain medium and air), as calculated from their Mie scattering spectral peaks.[15] For nanospheres with deep subwavelength diameters, the electric dipole (ED) and higher order modes (EQ to E32) converge within a narrow spectral range near 600 nm, corresponding to LSP resonances. As the diameter ($d$) exceeds the quasi-static limit, the resonance curves shift due to phase retardation effects.[16] Beyond this transition, the magnetic modes (MD to MO) develop, and both electric and magnetic modes follow linear dispersion curves that intersect the origin. Besides this LSP picture, the dispersion relationship can be explained from the perspective of SPP waves. Farther away from $\omega_{SP}$, the SPP wave refractive index becomes close to $n_m$. The asymptotic line for the ED corresponds to the half-wavelength Fabry-Perot (FP) resonance given by $\lambda \approx 2.81 n_m d$, EQ has a slope of $\lambda \approx 1.41 n_m d$, and so on.[17] Near $\omega_{SP}$, which in spheres correspond to the LSP resonant frequencies, the SPP wavelength is significantly reduced, causing the dispersion curves to bend (see Supplementary Note 4). In this view, LSP is the half-wave resonance of SPP near $\omega_{SP}$, and previous nano-lasers that utilized higher-order SPP waves can be viewed as the lasing of multipole LSP modes in the FP regime. An important insight gained from the dispersion curves is that while the ED in the quasi-static regime is spectrally close to other modes, leading to a large $M$, modal separation increases in the FP regime, allowing only the ED mode to fall within the wide bandwidth of a gain medium (Extended Data Fig. 1c).

We considered a pair of gold and III-V semiconductor nanodisks. Figure 1b depicts the Mie scattering spectra of a gold disk with a diameter ($D$) of 250 nm and a thickness of 100 nm, and a semiconductor disk with a refractive index of 3.5, matching diameter, and thickness of 130 nm. When the two disks are well separated, each exhibits its fundamental half-wave dipolar mode at 1.69 eV ($\lambda$ = 730 nm). As the two disks are brought closer, the modes are increasingly coupled, and the lowest-order hybrid ED mode emerges at 0.9 eV (1380 nm) upon contact. This fundamental mode localizes at the metal-semiconductor interface (Fig. 1c). The resonance peaks of hybrid modes for various disk dimensions are described in Supplementary Note 4. Our finite-difference time-domain (FDTD) analysis demonstrates that the strong mode coupling results in a cavity $Q$ of up to 20. Notably, $Q$ of ~10 is obtained when the gold disk is replaced by oversized plates (Extended Data Fig. 1d). This indicates that the edge of the semiconductor disk effectively provides both reflection (SPP picture) and charge localization (LSP picture).

In the following experiments, we utilized InGaAsP particles fabricated using a combination of dry and wet etching (Extended Fig. 2). By controlling the etching time per batch, we produced semiconductor particles with various sizes ranging from 100 to 300 nm and a thickness of 130 nm. These particles typically exhibited rhombus shapes. The particles were placed on a high-quality polycrystalline gold substrate (Fig. 2a).[18] We employed a 1064-nm pump laser (2.5 MHz repetition and 2 ns pulse width), along with a grating-based InGaAs-camera spectrometer (0.8 nm resolution) (Extended Fig. 3a). Figure 2c displays the emission spectra collected from 32 particles made of $In_{0.53}Ga_{0.47}As_{0.92}P_{0.08}$ at pump fluences of ~2 mJ/cm². The spectra are arranged along a theoretical tuning slope described by $\lambda \approx 3.09\,x + 596$ [nm], where $x$ represents the side-length of the rhombus. Particles smaller than 250 nm exhibit single emission peaks at wavelengths as short as 1200 nm (~ 370 nm lower than the band gap edge near 1570 nm). For larger particles, second peaks are observed, attributed to the ED modes along the shorter axes of the particles. We also conducted experiments with particles coated with a 5-nm thick insulating layer of silica and obtained similar results likely due to the surface roughness of InGaAsP particles.

As the pump fluence was varied, each particle showed nonlinear intensity growth, spectral narrowing, and single polarization peak over quasi-unpolarized background (Extended Data Fig. 3). For instance, at room temperature one device displayed a threshold at 350 µJ/cm² with a fitted $\beta$ factor of 0.06 (Fig. 2d), and an output spectrum at 1350 nm with a linewidth of 27 nm (a linewidth Q of 50) (Fig. 2e, Extended Data Fig. 3b). In FDTD, only one resonance mode is present within the broad fluorescence bandwidth. The electric field and plasmon charges localize at the gold-$SiO_2$ interface, generating far-field emission akin to that of a dipole situated near a metal surface (Fig. 2f-h). This output is efficiently collected in a vertical direction by an objective lens (~68% efficiency for a numerical aperture of 0.75)[19].

We compared particles on a gold substrate (plasmonic devices) with those on silica-coated silicon substrates (dielectric devices). A total 120 devices produced from 6 batches were examined, each with different average particle sizes and thickness of approximately 290 nm (see Extended Data Fig. 4). As the particle size increased, along with the oscillating mode order, the threshold showed a modest decrease (Fig. 2g), and the linewidth significantly reduced to 4-6 nm (resulting in a linewidth Q of 240-340). For dielectric devices, lasing was observed only for particle sizes larger than 880 nm, even at the highest pump fluence up to 8 mJ/cm$^2$.[18,20,21]

Built on a known semiconductor laser model (Extended Data Fig. 5), we formulated rate equations in the spectral domain (see Supplementary Note 3). The model did not explicitly consider expected spatial dependence in spontaneous and stimulated emission rates and the accompanying depletion and diffusion[22] of electron-hole pairs, treating them as spatially averaged parameters. Given the bang gap edge at 1570 nm and under optical pumping at 1064 nm, the total number of energy states available is estimated to be ~$1 \times 10^{19}\ cm^{-3}$. This large concentration of carriers is normally not reachable due to the Auger recombination (~$2.3 \times 10^{-29}\ cm^6/s$ for InGaAsP) but is attained in plasmonic cavities due to the enhanced radiative emission. Within the mode bandwidth ΔΩ (=ω/Q) centered at 1240 nm, the number of gain emitters is $3.4 \times 10^{18}\ cm^{-3}$ for $Q = 10$. In a bulk semiconductor, the theoretical maximum free-space gain would be around 3,000 cm$^{-1}$ at 1240 nm, which is insufficient to compensate for the cavity loss ($2n\pi/\lambda Q$) of ~17,700 $cm^{-3}$ (Supplementary Note 2). Within a subwavelength cavity, the cavity gain is enhanced by the Purcell factor $F_p$ (Extended Data Fig. 5e), making it possible to reach the threshold. Figure 3a illustrates a numerical result for a half-wave plasmonic laser far above threshold at a pump fluence of 0.8 mJ/cm$^2$.

Remarkably, our numerical simulation closely reproduced the experimental results. Figure 3b illustrates the measured and simulated spectra of two lowest-order (half-wavelength) devices and a one-wavelength device. At low pump power levels, only spontaneous emission near the band edge is evident. As the pump increases, the excited band fills up, and the spontaneous emission peak shifts toward higher energies. With stronger pumping, photons (polaritons) accumulate in the cavity, and eventually stimulated emission becomes greater than spontaneous emission. Upon reaching this threshold, a narrowing of the linewidth occurs, and the increase in the number of photons shows nonlinear growth (Extended Data Fig. 6). A kink, which is characteristic of $\beta \ll 1$, appears even when only one mode present in the gain bandwidth. This apparent kink in the light-in-light-out curve results from level filling; the amplification of the mode has to wait for the quasi-Fermi level of the emitters reaches the mode resonance. The so-called threshold-less lasing would not be possible unless the bandwidths of the mode and emitters perfectly overlap. The roll-off of the output power at higher pump levels can be attributed to increasing nonradiative Auger recombination. Ultimately, the output saturates when all excited states have been populated. When comparing the results at room temperature to those from Peltier-cooled samples, we observed a modest spectral narrowing (Extended Data Fig. 3d).

Isolated micro- and nano-lasers detached from substrates, also known as laser particles (LP), offer the potential for integration into various physical and biological systems, enabling applications that are not feasible with on-chip lasers. We prepared In$_{0.8}$Ga$_{0.2}$As$_{0.44}$P$_{0.56}$ disks or rhombuses on InP pillars, added a 5 nm-thick silica coating, and deposited a layer of gold through evaporation from the top (Fig. 4a). Electron microscopy of these plasmonic LPs revealed a conformal gold coating with a thickness of 80-100 nm, found exclusively on one side (Fig. 4b and Extended Data Fig. 7).

In contrast to the substrate-based devices explored earlier, optically pumped LPs with half-wavelength sizes fell short of reaching a lasing threshold. The smallest LP batch that exhibited lasing had a size of ~ 580 nm (Extended Data Fig. 8), corresponding to a longitudinal order of 3 and yielding a linewidth of 4 nm. We also observed narrowband lasing from LPs made with a gold thickness of 15 nm, thinner than the 30-40 nm skin depth of gold at these wavelengths (Extended Data Fig. 8). Enhancing the uniformity of the gold layer[23] may make lower-order LPs possible. The emission pattern from a half-wave LP with the thin, semi-transparent gold layer would closely resemble that of a point dipole (Fig. 4d). This diffraction-limit source might be aptly termed a 'laser dot,' akin to a quantum dot but with increased brightness and narrower spectra. LPs with a few-mode order still possess sizes smaller than the free-space wavelength and delivered linewidth Q factors exceeding 250. Figure 4e presents laser spectra from 31 LPs with varying diameters. The single-mode tuning range extends up to 150 nm and can be expanded using different III-V compositions.

For biocompatibility we functionalized LPs with polyethyleneimine coating. Following incubation with Hela cells, polymer-coated LPs were internalized into the cytoplasm. Upon optical pumping, these intracellular lasers emitted stimulated emission with a linewidth of 4 nm (Fig. 4g). The excitation energy of several picojoules per pump pulse is well-suited for biological applications, where the pump beam can be rapidly scanned across cells in imaging or in flow cytometry.[24]

In summary, our work has successfully demonstrated lasers operated in their fundamental oscillation frequencies by harnessing the half-wave dipolar mode. This achievement has resulted in the smallest device sizes when compared to previous higher-order devices, even surpassing those operating at cryogenic temperatures (see Extended Data Fig. 9). It is anticipated that proportionally smaller devices could be realized at shorter wavelengths down to 800 nm using other III-V, such as InP and GaAs. Furthermore, the substitution of gold with silver may lead to even narrower linewidths. However, it remains uncertain whether quasistatic-limit lasers, or their equivalent, sub-wavelength lasers near the surface plasmon frequencies in the visible, can be achieved using currently available material options. Crucially, our metal-semiconductor design has enabled us to produce submicron plasmonic laser particles capable of emitting narrowband light across a broad spectral range. These injectable light emitters hold promise for various applications including large-scale multiplexed imaging, single-cell analysis, and plasmonic sensing.[7–9,21]

**METHODS**

**Materials.** Custom semiconductor wafers with metal-organic chemical vapor deposition (MOCVD) epitaxial layers on InP substrates were ordered from Seen Semiconductors. Gold substrates were purchased from platypus and measured to have a surface roughness of 1.2 nm. $H_2SO_4$ was purchased from Transene Company. All other chemicals, such as $H_2O_2$, HCl, tetra-orthosilicate, polyethyleneimine, and ammonium hydroxide, were purchased from Sigma-Aldrich.

**Half-wave device fabrication.** We used an epitaxial wafer with three layers of undoped InGaAsP with different stoichiometry, separated by InP sacrificial layers. Mesa structures were fabricated using optical lithography and reactive ion etching to a diameter of 0.9 to 1 μm. Among the three InGaAsP materials, we used $In_{0.53}Ga_{0.47}As_{0.92}P_{0.08}$ layers with a pristine thickness of 290 nm. The diameter of the InGaAsP layer was varied by immersing the wafer chip in acid Piranha solution ($H_2SO_4$:$H_2O_2$:$H_2O$ = 1:1:10) etching for a pre-calibrated reaction time.[21] Then, the InP layers were removed using hydrochloric acid (HCl), which release InGaAsP particles.[7] The particles were washed with ethanol and water and dispersed in ethanol. The InGaAsP particles were added into Piranha solution ($H_2SO_4$:$H_2O_2$:$H_2O$ = 1:1:10) and etched for a pre-calibrated reaction time to the final, desired thickness as well as lateral size. Submicron particles typically acquired rhombus shapes due to anisotropic etching rates on different lattice planes of InGaAsP. For $SiO_2$ coating, we used Stöber method on the etched particles.[7] The particles were subsequently washed multiple times with ethanol and water and then drop-casted onto designated gold or silica-coated silicon substrates.

**Optical experiments.** We utilized a commercial laser-scanning microscope platform (Olympus, FVMPE-RS). The laser-scanning unit of the microscope was coupled with both nanosecond and picosecond pump lasers. The nanosecond pump laser (IPG photonics, YLPN-1-1X120-50-M) at 1064 nm had a tunable pulse duration ranging from 1 to 120 ns and a variable repetition rate from 2 kHz to 14 MHz. For shorter pump pulses, we used a picosecond pump laser (Picoquant, VisIR-765) with a pulse duration of 70 ps, a center wavelength of 765 nm, and a repetition rate variable from 31 kHz to 80 MHz. A near-infrared-optimized, 100x, 0.85-NA (numerical aperture) objective (Olympus IMS LCPLN100xIR) was used for optical pumping. The output emission from samples was collected using the same lens and directed to a near-infrared spectrometer via a dichroic mirror. An InGaAs linescan camera (Sensor Unlimited 2048L) was used for spectra characterization with a typical integration time of 1 ms. The spectrometer had a diffraction grating with 200 lines per millimeter, providing a resolution of 0.8 nm and a wavelength span of 1100 to 1600 nm. For low-temperature experiments, we placed samples on a peltier cooler connected to a chiller.

**Numerical Simulations.** The FDTD calculations of Mie scattering spectra and dipole-embedded cavity mode properties were performed using a commercial software (Lumerical FDTD Solutions). More details are available in Supplementary Note 5.

**Semiconductor laser modeling.** The rate equations of a semiconductor laser were formulated and solved numerically using MATLAB in a laptop. Briefly, the energy states of electron-hole pairs ($q$) are divided into typically 120 levels (labeled 1 to k) and the numbers of emitters evolve over time via pumping ($P_k$), stimulated emission ($q * \tau_s^{-1}$), spontaneous emission ($\tau_s^{-1}$), and non-radiative decays to lower levels ($\tau_{nr}^{-1}$). The photon numbers with energies labeled from 1 to k are varied over time through stimulated emission, spontaneous emission, and cavity loss $\tau_c^{-1}$. Thermodynamic excitation was neglected for computational simplicity.

For electron-hole pair numbers in the 1st (band edge) to the $k$-th (pump) levels:

$$\frac{dn_k}{dt} = P_k(t) - \frac{n_k}{\tau_s(\omega_k)} - \frac{n_k}{\tau_{nr}(\omega_k)}$$

$$\frac{dn_{k-1}}{dt} = -\frac{\Delta\Omega}{\delta\omega}\frac{n_{k-1}q_{k-1}}{\tau_s(\omega_{k-1})} - \frac{n_{k-1}}{\tau_s(\omega_{k-1})} - \frac{n_{k-1}}{\tau_{nr}(\omega_{k-1})}$$

$$\frac{dn_{k-2}}{dt} = -\frac{\Delta\Omega}{\delta\omega}\frac{n_{k-2}q_{k-2}}{\tau_s(\omega_{k-2})} - \frac{n_{k-2}}{\tau_s(\omega_{k-2})} - \frac{n_{k-2}}{\tau_{nr}(\omega_{k-2})} \quad ...$$

$$\frac{dn_1}{dt} = -\frac{\Delta\Omega}{\delta\omega}\frac{n_1 q_1}{\tau_s(\omega_1)} - \frac{n_1}{\tau_s(\omega_1)}$$

For photons (polaritons) in the $i$-th levels ($i = [1: k-1]$):

$$\frac{dq_i}{dt} = \frac{\Delta\Omega}{\delta\omega}\frac{n_i q_i}{\tau_s(\omega_i)} + \frac{n_i}{\tau_s(\omega_i)} - \frac{q_i}{\tau_c(\omega_i)}$$

The transition rate coefficients depend on the resonance frequency $\omega_o$ and bandwidth $\Delta\Omega$:

$$\frac{1}{\tau_s(\omega_i)} = \frac{1}{\tau_s} \cdot \frac{\Delta\Omega^2/4}{(\omega_i - \omega_o)^2 + \Delta\Omega^2/4}$$

$$\frac{1}{\tau_c(\omega_i)} = \frac{1}{\tau_{c0}} \bigg/ \frac{\Delta\Omega^2/4}{(\omega_i - \omega_o)^2 + \Delta\Omega^2/4}$$

$$\frac{1}{\tau_{nr}(\omega_i)} = \frac{1}{\tau_{nr0}} \cdot \frac{n_{sat}(\omega_{i-1}) - n_{i-1}}{n_0}$$

Details are described in Supplementary Note 4.

**Fabrication of metal-coated semiconductor particles.** To produce stand-alone laser particles, we used a $In_{0.73}Ga_{0.27}As_{0.58}P_{0.42}$ single-layer wafer. After optical lithography and RIE etching, pillars with a diameter of ~ 2 µm was created.[7] The diameter of the InGaAsP layer was reduced by immersing the wafer chip into acid Piranha solution ($H_2SO_4:H_2O_2:H_2O$=1:1:10) for a specific etching duration.[21] Subsequent HCl or $H_3PO_4$ was used to etch InP sacrificial layers to diameters roughly matching that of InGaAsP layer as well as removing an InP cap layer on top the InGaAsP layer. After washing the wafer, a 5-nm thick layer of $SiO_2$ coating was produced on the pillars using the Stöber method.[7] A gold layer was deposited on the pillar array using an e-beam evaporator (Denton EE-4, Harvard Center for Nanoscale Systems). By immersing the wafer in HCl, gold-coated metal-semiconductor particles were detached from the substrate. The harvested particles were washed with fresh water and ethanol three times. To enhance material stability and biocompatibility in aqueous media and the cytoplasm, approximately 60 nm thick silica shell was added to the particles by the Stöber method in three cycles.[7] To further enhance biocompatibility and facilitate cellular uptake, the silica-coated metal-semiconductor particles were encapsulated with polyethyleneimine (PEI) polymers.

**Cell experiments.** Green fluorescent protein (GFP) expressing HeLa human cervical cancer cells were purchased from GenTarget. The HeLa cells were cultured in Dulbecco's modified Eagle medium (DMEM) supplemented with 10% (v/v) fetal bovine serum (FBS) and 1% (v/v) penicillin–streptomycin, at 37 °C under 5% $CO_2$. For cell tagging, cells of a desired density were plated on a glass-bottom plate with the cell media. Metal-semiconductor laser particles suspended in the same cell culture media were added to the cell containing plate. The plate was shaken to disperse the laser particles evenly. After 24 h of further incubation, cell laser experiments were performed using a customized microscope.

**Acknowledgements**

The Yun lab laser project team (Drs. Yue Wu, Debarghya Sarkar, and Kwonhyeon Kim) is acknowledged for helpful discussions. This study was supported by National Institutes of Health research grants (R01-EB033155, R01-EB034687). S.C. acknowledges the MGH Fund for Medical Discovery fundamental research fellowship award. This research used the resources of the Center for Nanoscale Systems, part of Harvard University, a member of the National Nanotechnology Coordinated Infrastructure, supported by the National Science Foundation under award number 1541959.

**Author contributions**

S.C. and S.H.Y. designed the study. S.C. performed experiments and FDTD simulation. N.M. contributed to the optical setups. S.H.Y. conducted semiconductor laser modeling. S.C. and S.H.Y. analyzed the data and wrote the manuscript.

**Financial Conflict of Interest**

N.M. and S.H.Y. have financial interests in LASE Innovation Inc., a company focused on commercializing technologies based on laser particles. The financial interests of N.M. and S.H.Y. were reviewed and are managed by Mass General Brigham in accordance with their conflict-of-interest policies.

**Data availability**

The laser experiment data used within this paper are available at https://doi.org/10.7910/DVN/AOWZN8 (ref.[25]). The authors can provide the additional data that supports the findings of this study upon a reasonable request.

**Code availability**

The semiconductor laser modeling and laser experiment data analysis code used within this paper are available at https://doi.org/10.7910/DVN/AOWZN8 (ref.[25]). The details of FDTD simulations and semiconductor laser modeling can be found in the Methods and Supplementary Information. The associated codes can be obtained from the authors upon a reasonable request.

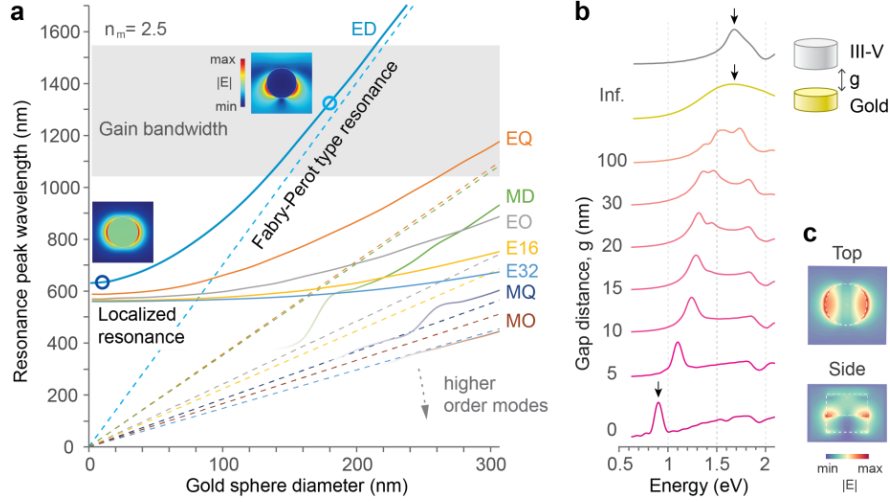

**Fig. 1. Strong coupling of plasmonic and semiconductor dipolar modes. a**, Theory for the peak wavelengths for different resonance modes (ED: Electric dipole, EQ: Electric quadrupole, MD: Magnetic dipole, EO: Electric octupole) at various diameter of the gold sphere in a surrounding index of 2.5. Solid curves are for gold, and dashed curves are for a perfect conductor with an infinite plasma frequency. Insets, the electric ($|E|$) profiles of the ED modes in the quasi-static and dynamic regimes, respectively. In the FP regime, only the ED mode can be placed within the gain bandwidth while all the higher order modes excluded. **b**, Simulated Mie scattering spectra of a complex of semiconductor ($n$=3.5) and gold nanodisks with the same diameter of 250 nm for different gap distances. Arrows indicate the locations of the ED modes. **c**, Electric field amplitude profiles of the semiconductor-gold structure in contact (dashed outline).

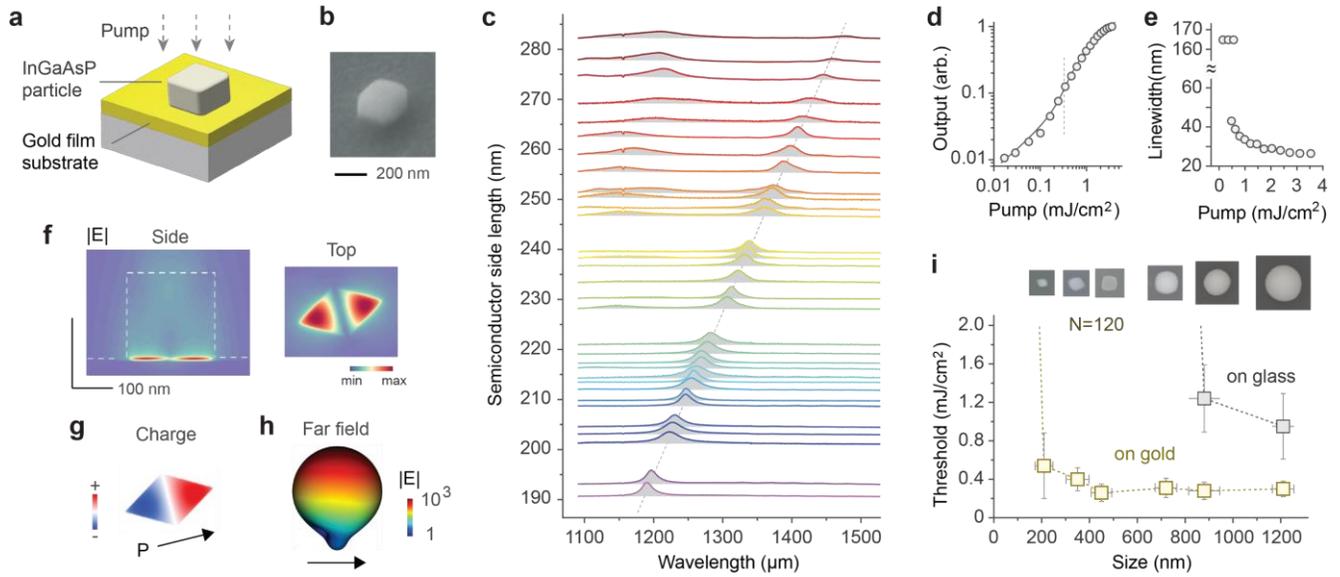

**Fig. 2. Characteristics of semiconductor-on-gold nanodevices. a**, Schematic of the semiconductor-on-gold design. **b**, SEM image of an InGaAsP particle with a side length of 170 nm (280 nm from corner to corner). **c**, Emission spectra from particles with different sizes. The dotted line indicates FDTD-calculated tuning curve. **d**, Measured light-light curve (circles) of a sample and theoretical fit (solid curve, $\beta$=0.06). **e**, Measured linewidth at different pump fluences. **f**, Field amplitude ($|E|$). **g**, Induced charges ($\nabla \cdot E$). **h**, Computed far-field pattern of a half-wave device. **i**, Measured laser thresholds of several sample batches fabricated for different semiconductor sizes. Inset, SEM images of representative samples from different batches.

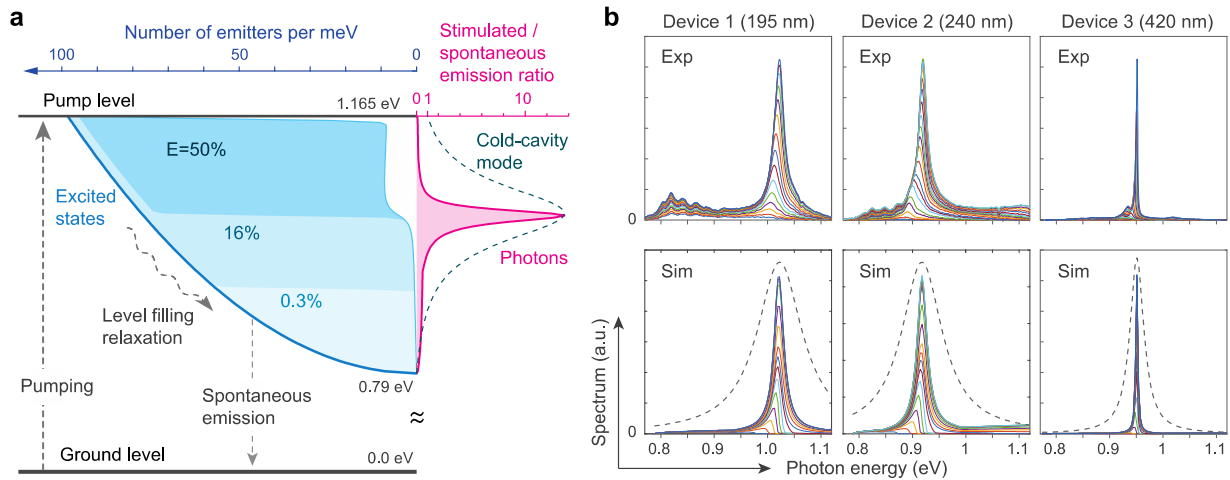

**Fig. 3. A semiconductor laser 'waterfall' model. a**, Energy diagram of gain emitters (electron-hole pairs) and photons (polaritons) in a cavity. The population in the excited states (bluish shades) and output spectra (cyan) are illustrated for a 240 nm-size device at a pump fluence of 0.8 mJ/cm². In the beginning of a pump pulse (0.3% of the total pump energy), only the bottom excited states are filled. As the pump increases (to 16%), the quasi-Fermi level elevates to the mode center frequency, at which the threshold has been reached. At the peak of the pump pulse (50% energy), stimulated emission becomes greater than the spontaneous emission of the mode. See Extended Data Figure 5 and Supplementary Note 4 for more details. **b**, Comparison of simulation and experiment for two lowest-order (half-wave) devices (simulated Q = 10) and one high-order device (simulated Q = 30), for a range of pump fluence from 0.03 to 3 mJ/cm². A periodic spectral fringe appears between 0.8 and 0.88 eV owing to internal interference in a dichroic mirror in the setup. See Extended Data Figure 6 for light-in-light-out curves and linewidths.

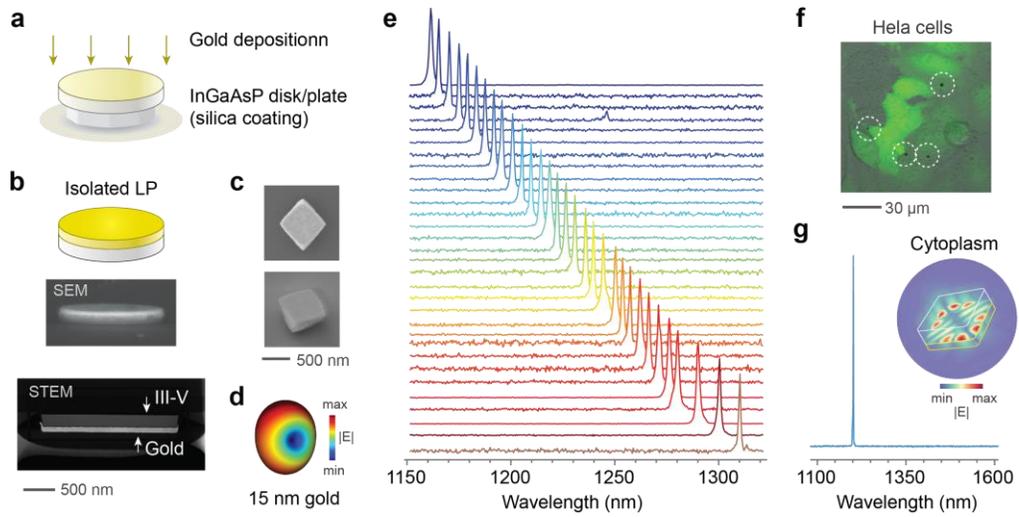

**Fig. 4. Plasmonic laser particles and intracellular lasing. a**, Schematic of fabrication. **b**, Electron micrographs of gold-semiconductor LPs. **c**, 580-nm sized samples. **d**, Calculated far-field pattern of a dipolar LP with a 15 nm thick semi-transparent gold. **e**, Output spectra of 31 LPs over a wide wavelength range with pumping at ~20 pJ/μm² per pulse. **f**, Hela cells (expressing green fluorescent protein) tagged with LPs (circles). **g**, Measured spectrum and FDTD simulation of an intracellular metal-semiconductor laser. Scale bar, 500 nm.

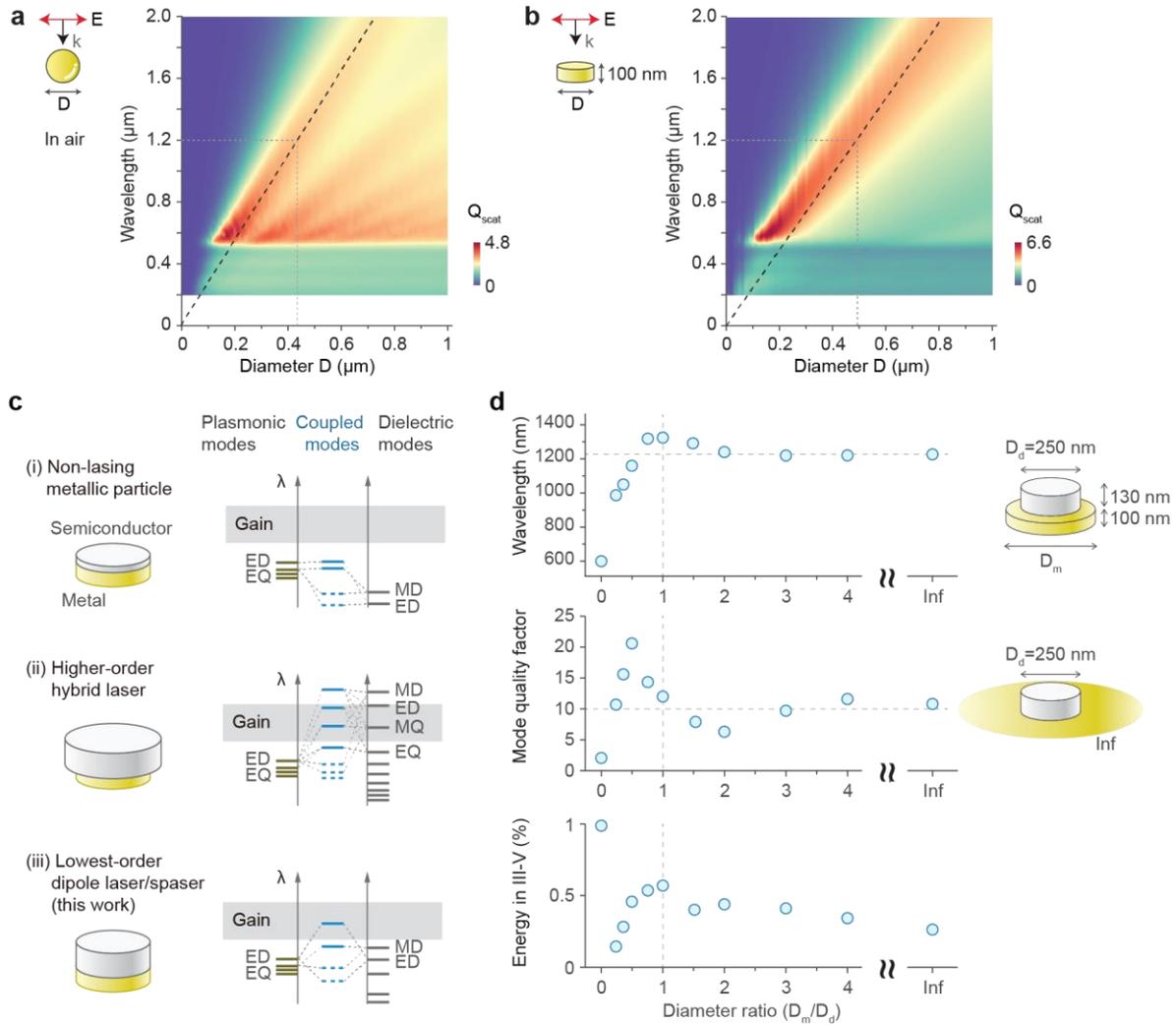

**Extended Data Fig. 1. Mode properties of metal-semiconductor nanoparticles**. **a-b**, Mie scattering spectra of gold nano-spheres (a) and gold nano-disks (b) in the air for planar incident waves. While the excitation of higher order modes is evident for spheres, only the fundamental electrical dipole mode clearly appears for disks owing to the symmetry; The higher modes in disks are not efficiently excited by the uniform driving field. In nano-lasers, however, the higher-order plasmonic modes are driven by local emitters and can be efficiently excited via near-field interactions. **c**, Schematic depicting mode coupling between plasmonic and semiconductor disk modes for three representative cases: (i) Non-lasing metallic luminescence when a semiconductor disk is too thin. Because of the large differences of the modes in energy, mode coupling is weak, and the lowest order modes are largely plasmonic. Because of the proximity of the modes, it is difficult to selectively amplify only the ED mode; (ii) Higher-order hybrid laser, where multiple dielectric-like modes are present within a gain bandwidth; (iii) A hybrid dipole laser — this may be regarded a "spaser" — where the individual modes in the metal and semiconductor disks have similar energies. Strong coupling occurs between ED modes, separating the hybrid plasmon-like mode from the other hybrid modes. This mode shift may be considered as the effect of the refractive index of the semiconductor on the plasmonic mode. However, mode coupling is a more accurate explanation as the effective index the plasmonic ED mode experiences is the same as the index of the ED mode in the dielectric medium. Note that the MD modes, the lowest order modes in dielectric disks, are not efficiently coupled with the plasmonic ED mode because of the field symmetry. **d**, FDTD simulation of metal-semiconductor disks with different diameters depicted in the inset. The resonance wavelength, quality factor, and mode confinement factor in the semiconductor vary as a function of the diameter ratio from 0 (III-V only) to infinity (on a gold substrate). The resonance wavelength of the hybrid ED mode increases dramatically from 600 nm to 1300 nm at size matching and then 1220 nm for oversized gold. The quality factor has the maximum at a diameter ratio of 0.5, in part due to the best mode energy matching and in part due to the lower metallic absorption at 800-900 nm. The Q factor approaches to slightly over 10 at infinite gold, where a quarter of electromagnetic energy residing in the semiconductor while the rest three quarters are stored in the metal. See Supplementary Figure S2 for more examples of Mie scattering spectra.

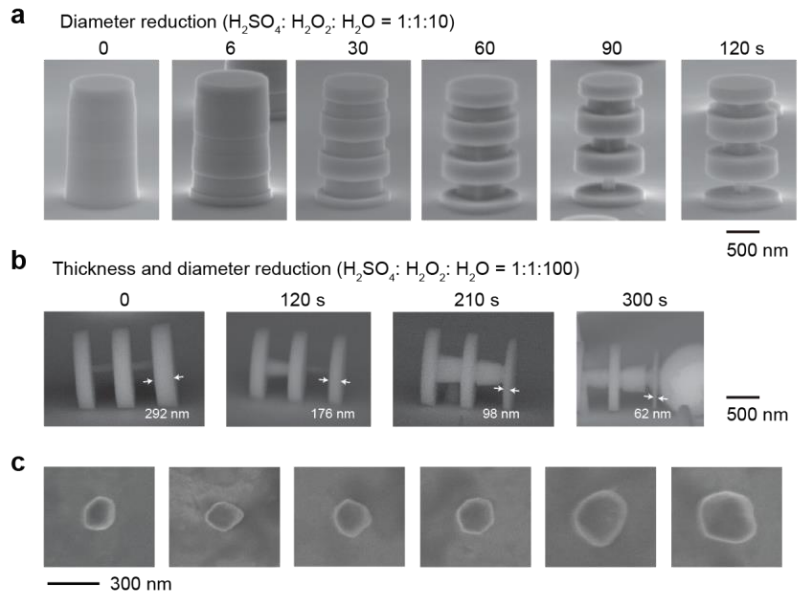

**Extended Data Fig. 2. Size and thickness tuning of semiconductor particles. a**, Two-dimensional etching for reducing the diameters of InGaAsP layers while preserving their thickness using piranha acid solution. The InP layer remains the same size while InGaAsP is etched away. The bottom layer with a composition of $In_{0.53}Ga_{0.47}As_{0.92}P_{0.08}$ was primarily used in most experiments (unless specified). **b**, Three-dimensional etching of InGaAsP layers, performed after a full (typically) or partial (for this dataset) etching of InP layers between InGaAsP layers. This process reduces both the thickness and lateral sizes. The 2D and 3D etching techniques were judiciously used to obtain desired thicknesses and sizes for InGaAsP particles. **c**, SEM images of six particles obtained from a single batch targeting a thickness of 130 nm and a mean side length of 250 nm. This batch was used to produce the experimental data in Figure 2c.

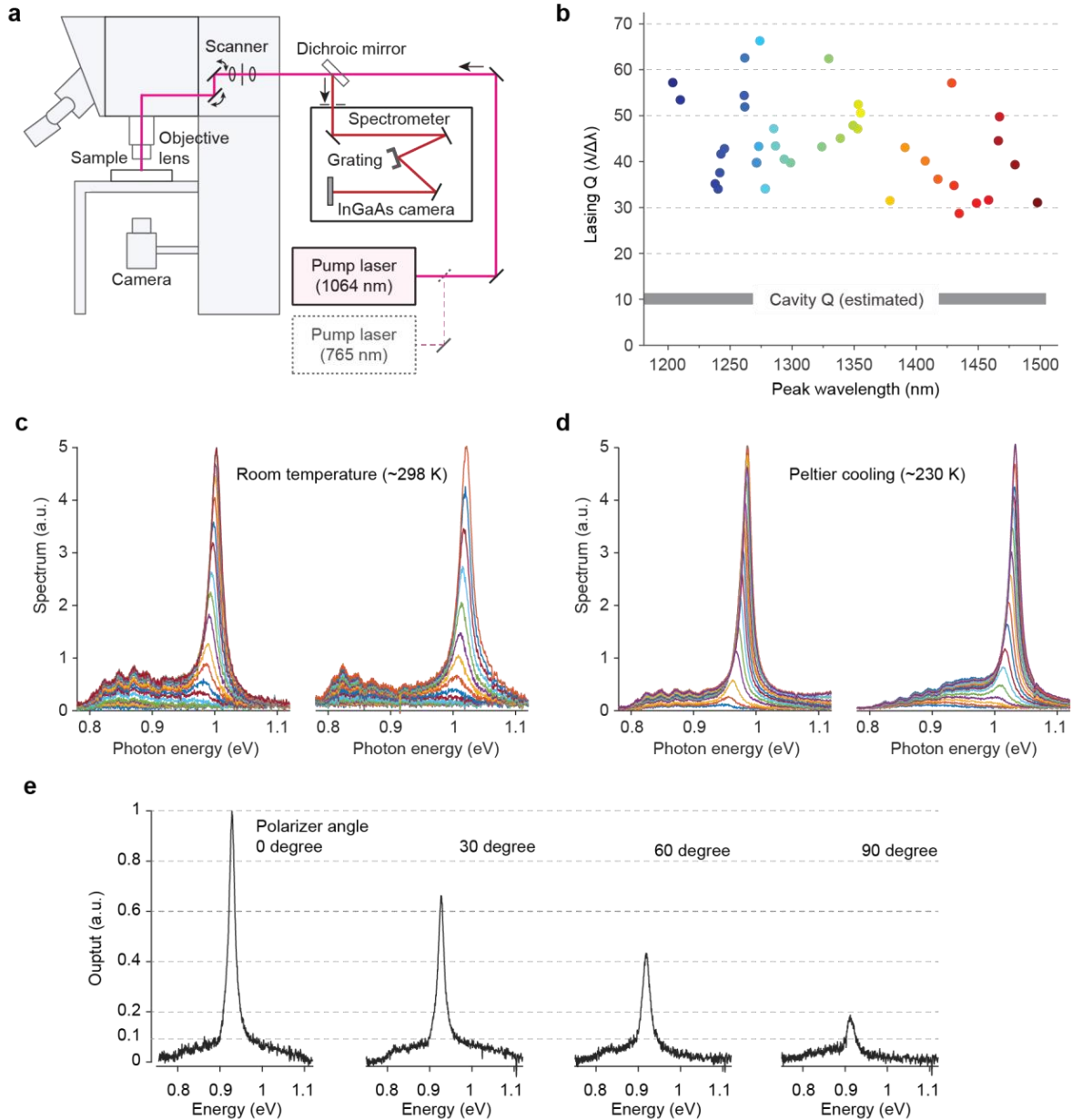

**Extended Data Fig 3. Half-wave dipolar lasers. a**, Schematic of a microscope setup used for optical characterizations. **b**, Measured lasing linewidth Q factors of 40 devices with different sizes and shapes. Representative spectra are displayed in Figure 2c. **c**, Emission spectra of two devices at a room temperature of 298 K. **d**, Two devices at a Peltier cooled temperature of ~ 230 K (nominal). Compared to the room temperature spectra, the falling edges at the high energy side, or near the quasi-Fermi levels, are steeper, presumably due to slightly reduced thermal excitations at the lower temperature. Note that the computer simulation spectra exhibit ever steeper spectral falloff at the quasi-Fermi level (see Fig. 3b and Extended Data Fig. 6), because no thermal excitations have been considered in the model, which corresponds to zero-degree temperature (0 K). **e**, Output

spectra through a polarizer at different angles. The stimulated emission peak at 0.94 eV (1321 nm) is linear polarized while the broad lower-energy background above the peak (0.75 to 0.9 eV) is approximately unpolarized.

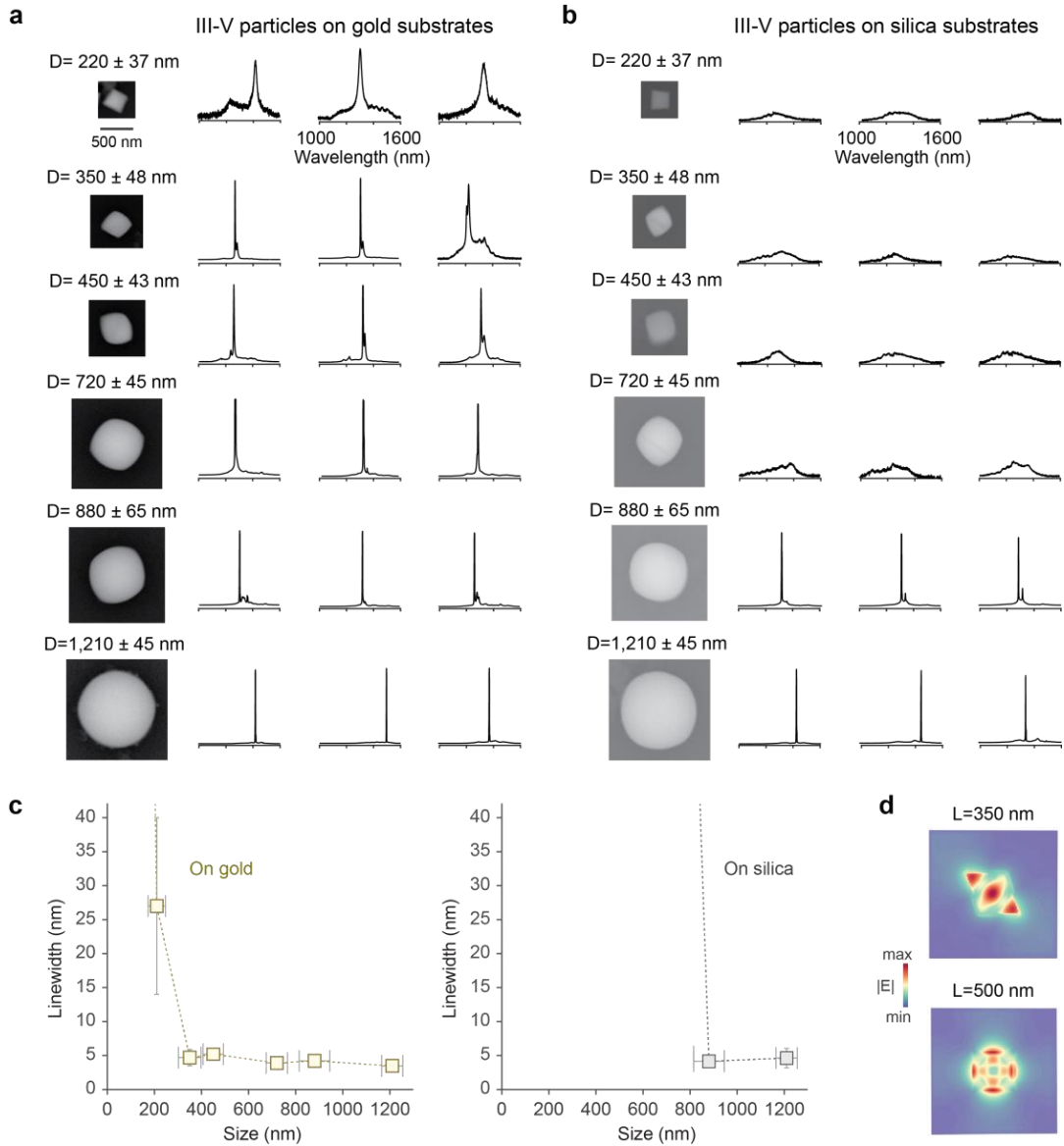

**Extended Data Fig 4. Higher-order mode devices. a-b**, Representative devices from different batches of varying sizes (same thickness of 290 nm) for plasmonic (**a**) and dielectric cavities (**b**). Dielectric devices with less than 880 nm sizes did not reach lasing threshold even at the highest pump power levels. **c**, Emission linewidths measured from a total 120 devices. The threshold pump fluences of these devices are shown in Figure 2is. **d**, FDTD results of a 350-nm rhombus-shape showing a second order mode at 1206 nm (top) and a rectangular-shape 500-nm device showing a whispering gallery mode (Q = 94) at 1164 nm (bottom).

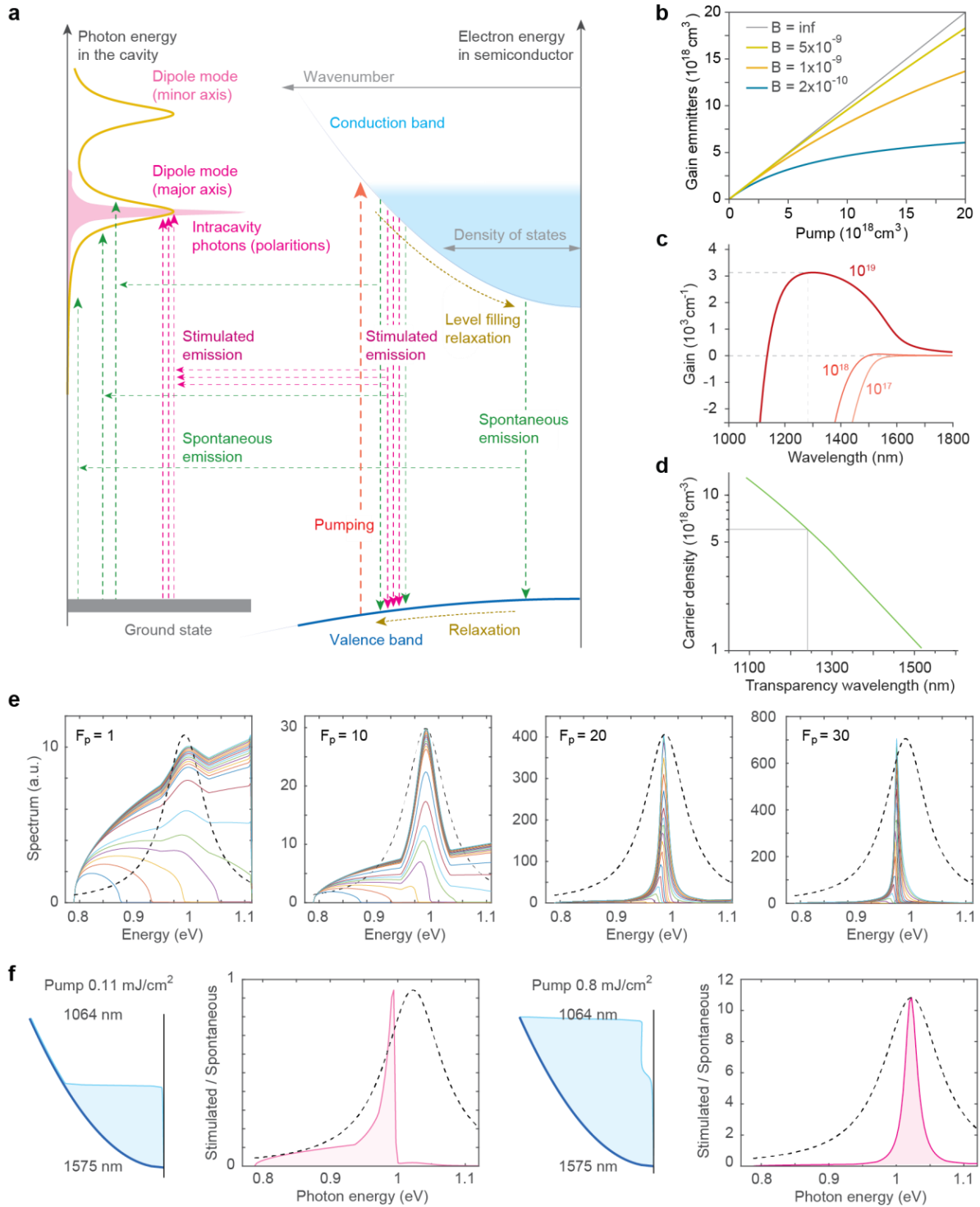

**Extended Data Fig 5. Semiconductor gain and a 'waterfall' laser model. a**, Energy level diagram and various transitions paths in a semiconductor laser. This forms essentially a four-level laser system (or a quasi-three-level including valence band absorption of intracavity light). The blue shade represents free electrons (or the electron-hole plasma) that fill the electronic states in the conduction band. The simplified "waterfall" model depicted in Fig,

3a is based on this diagram. **b**, Analysis of charge carrier loss due to Auger recombination for bulk (blue) and Purcell-enhanced (yellow) radiative decays. See Supplementary Note 2. **c**, Gain profiles at room temperature at three different carrier density levels, calculated using standard semiconductor theory considering the Fermi-Dirac distribution of the carriers at room temperature. The thermodynamic excitation was neglected in our numerical modeling in this work, which makes a sharp gain cliff beyond the quasi-Fermi level. **d**, Calculated total carrier density versus transparency (zero-gain) wavelength. **e**, Simulated output spectra of a device with a size of 240 nm for the cases of different Purcell factors, from 1 to 30, as the pump fluence is varied from 0.021 to 2.1 mJ/cm². The output saturates. Dashed curve illustrates the cold-cavity mode profile with a Q factor of 10. At $F_p = 1$, lasing threshold is never reached even at extreme pumping. At $F_p = 10$, lasing threshold is barely reached with a stimulated-to-spontaneous ratio of 1.07. Compared to $F_p = 20$, $F_p = 30$ results in reduced linewidths. Best correspondence to experimental data was obtained with $F_p = 18$ (Device 1) and $F_p = 19$ (Device 2). **f**, (Left) The emitter population and stimulated-to-spontaneous ratio at a threshold pump fluence, at which the quasi-Fermi level is just below the modal resonant frequency. (Right) At a pump level 7.3 times above the threshold when all the entire excited states are almost filled.

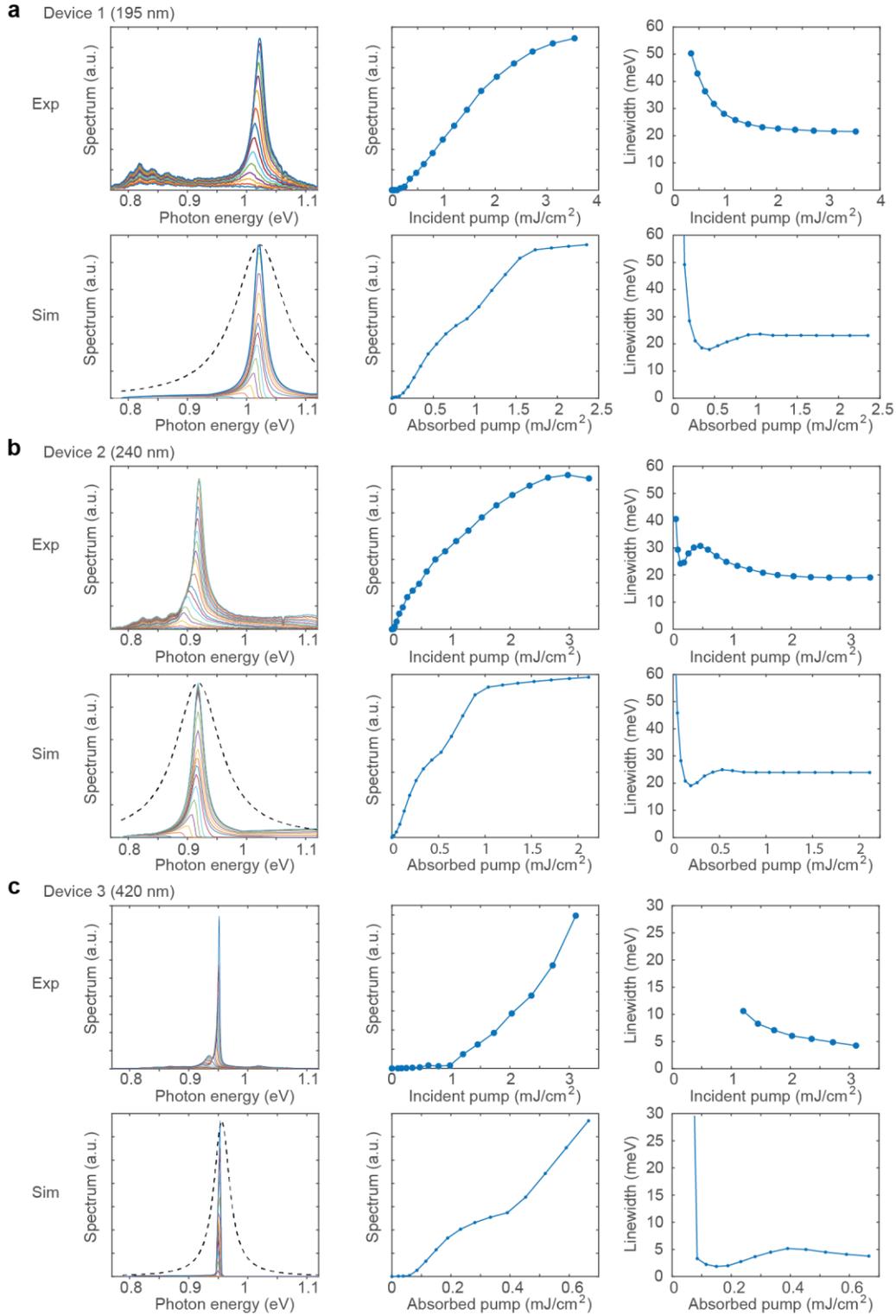

**Extended Data Fig 6. Laser simulation and experimental results. a**, Half-wave device (semiconductor volume: 183 x 183 x 130 nm$^3$). **b**, Half-wave device (225 x 225 x 130 nm$^3$). **c**, One-wave device (400 x 400 x 130 nm$^3$).

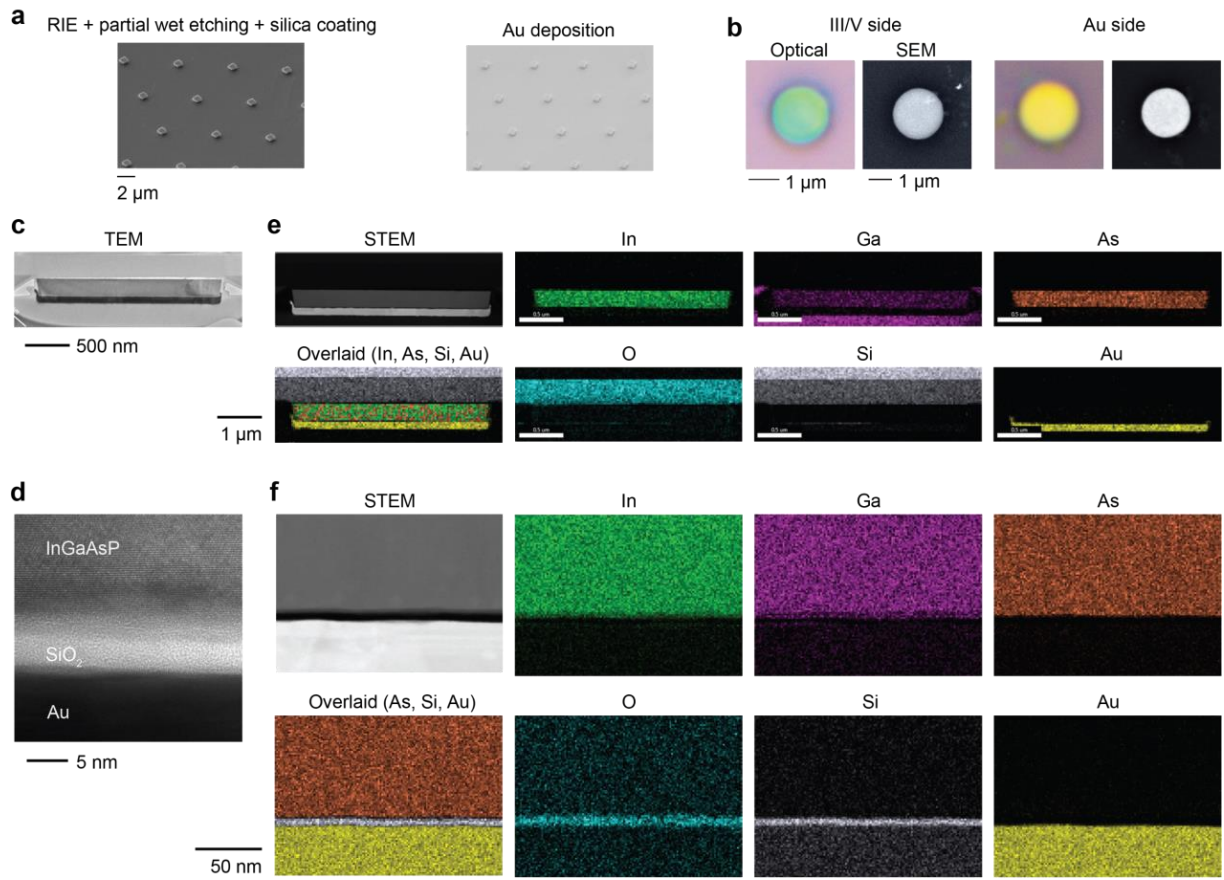

**Extended Data Fig 7. Structure of metal-semiconductor particles**. **a,** Schematic and SEM images of disk-on-pillar array following SiO$_2$ coating and gold deposition. **b**, Bright-field and SEM images of plasmonic laser particles drop casted on a silica-coated silicon substrate, presenting both the III/V side (left) and the Au side (right). **c,** Transmission Electron Microscopy (TEM) image of a cross-section of a sample, prepared by using focused ion-beam (FIB) etching. The image reveals the InGaAsP layer with a thickness of ~290 nm, 5-7 nm thick SiO$_2$ layer, and ~ 80-nm thick gold layer. **d**, Higher magnification view of the cross section. **e-f**, Scanning transmission electron microscopy (STEM) images and elemental maps. The samples in **c-f** were prepared by placing the InGaAsP side on a silica-coated silicon substrate and deposited Ga on top of the Au side of the particles. The samples were then placed up-side down on a TEM grid.

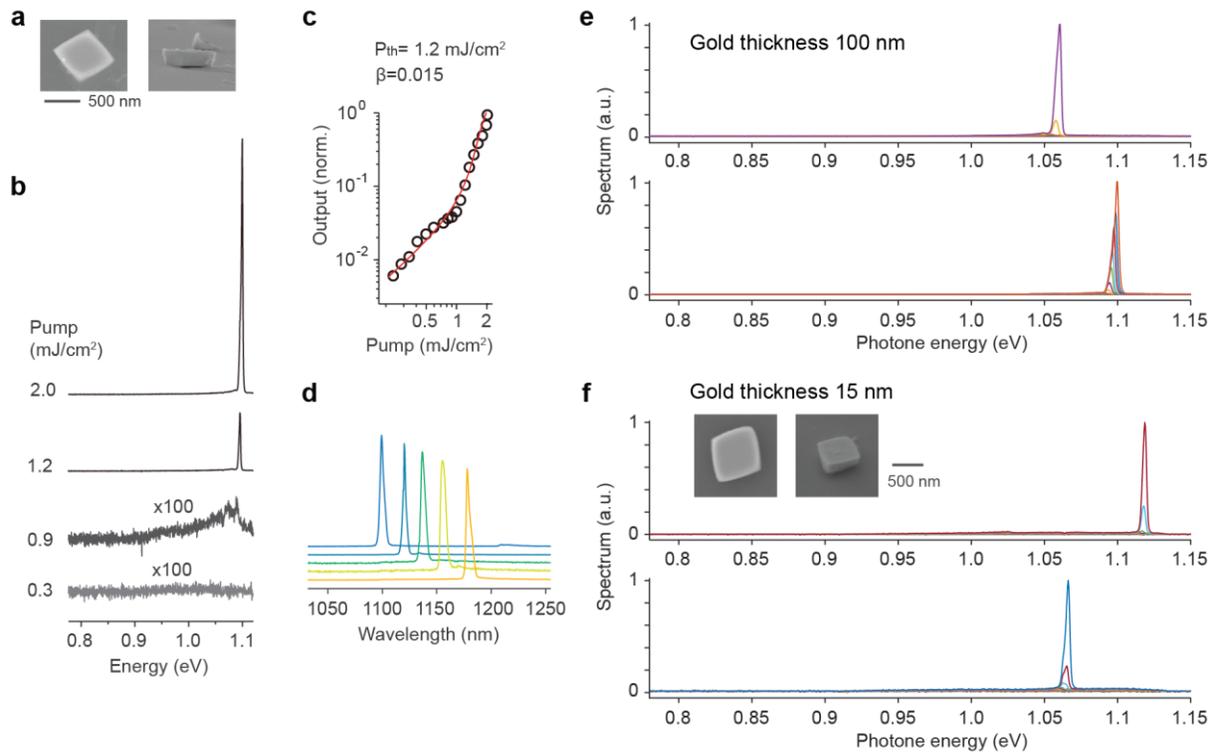

**Extended Data Fig 8. Emission spectra of isolated plasmonic LPs. a**, SEM images of samples with a gold thickness of 100 nm and a side length of 580 nm. **b**, Emission spectra at varying pump fluences. **c**, Light-in-light-out curves measured (circles) along with a theoretical fit (red curve). **d**, Output spectral of 4 LPs. **e**, Evolution of output spectra of two LPs at pump fluences varying from 0.1 to 4 mJ/cm$^2$. The Q factor of the fourth order longitudinal mode in these devices is estimated to be 42. **f**, LPs made with a gold thickness of 15 nm. Out spectra of two samples. Inset, SEM of samples with gold layers facing up. The estimated Q factor of the fourth order mode in these devices is 30.

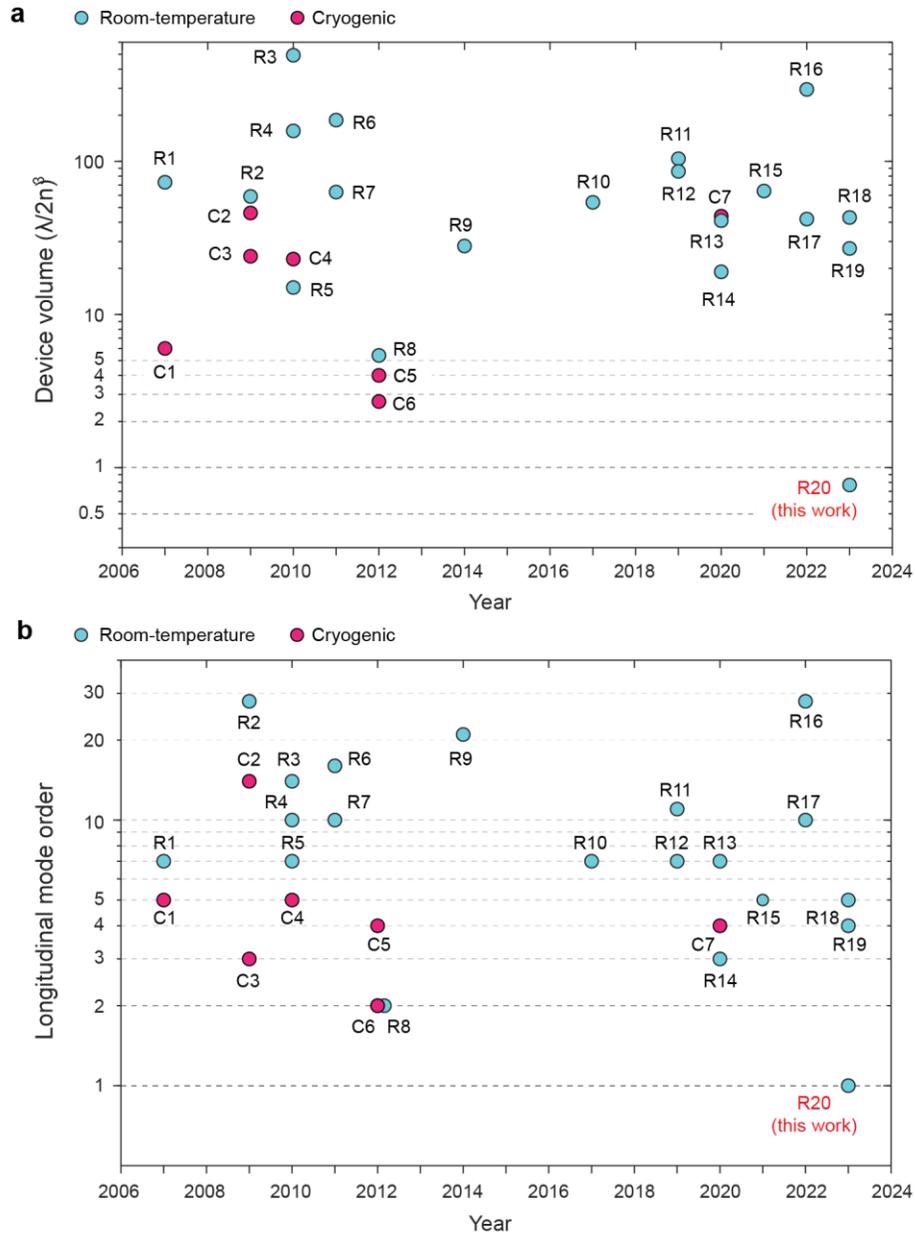

**Extended Data Fig 9.** Device volume (top) and the longest dimension (bottom) of single particle and micro- and nano-lasers operating at room temperature (cyan circles) and cryogenic temperature (pink circles). Each data point is labeled with the device name used in Supplementary Tables 1 and 2.

Supplementary Information for

# Half-Wave Dipolar Metal-Semiconductor Laser

Sangyeon Cho[1,2], Nicola Martino[1,2], Seok-Hyun Yun[1-3]*

[1]Wellman Center for Photomedicine, Massachusetts General Hospital, 65 Landsdowne St., Cambridge, MA 02139, USA

[2]Harvard Medical School, Boston, MA 02139, USA

[3]Harvard-MIT Health Sciences and Technology, Cambridge, MA 02139, USA

*Correspondence to: syun@hms.harvard.edu

## Supplementary Note 1. A simple laser model and the lasing threshold conditions

*1.1 Lasing threshold and the minimum gain emitter density*

Figure S1a depicts a simple model for semiconductor lasers. Upon the absorption of a pump photon, one free electron is excited to the pump level and then rapidly relaxed to the upper state near the bottom of the conduction band. The total number of electrons accumulated in the upper state is denoted $N_1$. The electron is transitioned to the lower state in the valence band, from which it further relaxes to the ground state. This forms a four-level laser system. The cavity can support more than one optical (plasmonic) mode within the gain bandwidth of the emitters. We assume that each of the emitters (electron-hole pairs in the semiconductor gain medium) interact with a total of $M$ cavity modes with significant coupling efficiency. For simplicity, we first ignore spatial dependence, assuming that each emitter overlaps equally with the $M$ modes and has the same transition efficiency represented by $\frac{1}{\tau_s}$ per mode. The spontaneous emission factor $\beta$ is then equal to $1/M$.

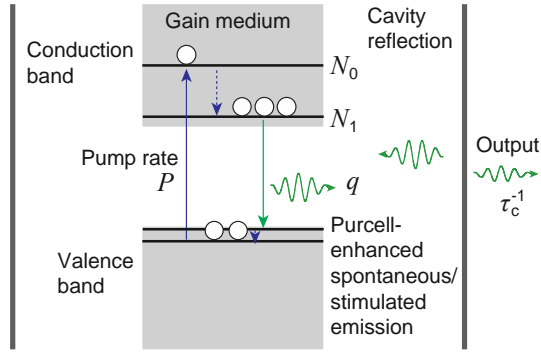

**Figure S1a.** Schematic of the laser model.

The laser rate equations for a specific lasing mode can be written as follows:

$$\frac{dN_1(t)}{dt} = P(t) - \frac{N_1 q}{\tau_s} - M\frac{N_1}{\tau_s} \quad (1)$$

$$\frac{dq(t)}{dt} = \frac{N_1 q}{\tau_s} + \frac{N_1}{\tau_s} - \frac{q}{\tau_c} \quad (2)$$

Here, $N_1$ denotes the number of the excited gain elements, $q$ the number of photons in the particular lasing mode of interest, $P$ is he pump rate, $\tau_s$ is the reciprocal of the radiative spontaneous emission rate into the mode, and $\tau_c$ is the lifetime of the mode in the cavity. The cavity Q-factor is defined as $Q = \omega\tau_c$, where $\omega$ is the optical angular frequency of the mode. The term $\frac{N_1 q}{\tau_s}$ describes optical amplification or the stimulated emission of $N_1$ emitters into the laser mode. The steady state solutions are:

$$q = P_{th}\tau_c/2 \left\{ \frac{P}{P_{th}} - 1 + \sqrt{\left(\frac{P}{P_{th}} - 1\right)^2 + \frac{4}{M}\left(\frac{P}{P_{th}}\right)} \right\} \quad (3)$$

where

$$P_{th} = \frac{M}{\tau_c} \quad (4)$$

From Eq. (2) we get:

$$N_1 = \frac{\tau_s}{\tau_c} \frac{q}{1+q} \quad (5)$$

These solutions are plotted as a function of $p$ in Figure S1b. From the graph, it is apparent that $P_{th}$ can be considered a threshold pump rate. At $P = P_{th}$, we get $q_{th} = \sqrt{P_{th}\tau_c} = \sqrt{M}$, and $N_{th} = \frac{\tau_s}{\tau_c} \frac{\sqrt{M}}{1+\sqrt{M}}$, which represents the minimum number of emitters to reach the threshold. $N_{th}$ is nearly invariant with the number of modes $M$ because the other non-lasing modes increase the total spontaneous emission rate and allow the emitters to be recycled more rapidly. Nonetheless, these modes increase the required pump energy by a factor of $M$. the pumping process always involves nonradiative energy transfer to phonons, resulting in heating of the gain medium. Moreover, for plasmonic cavities, the spontaneous emission of all the modes lead to heating of the metal. Since heating is often the performance limiting factor in low Q nano-lasers, it is highly beneficial to reduce $M$.

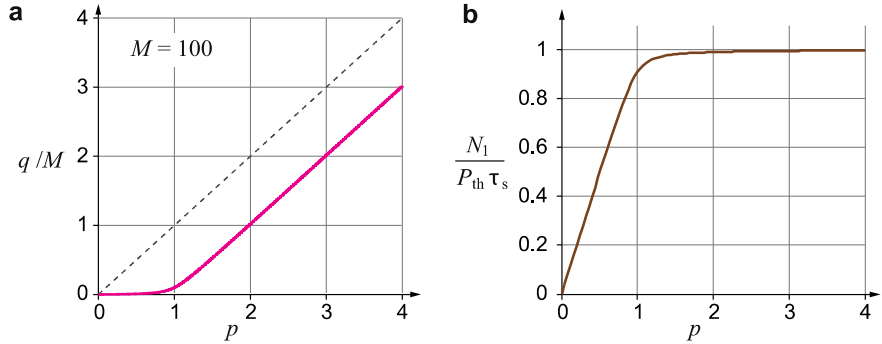

**Figure S1b.** Numerical calculation of the steady-state parameters for $M = 100$.

The minimum emitter density is $\rho_{min} = \frac{N_{th}}{V}$, where $V$ denotes the light-matter interaction volume. The per-mode radiative transition rate of $\frac{1}{\tau_s}$ can be related to the spontaneous emission time, $\tau_0$, of the emitters in free space (outside the cavity) as a reference independent of the laser design. By the definition of a modal

Purcell factor $F_p$, we can write $\frac{1}{\tau_s} = \frac{F_p}{\tau_0}$, Inserting $F_p = \frac{6}{\pi^2}\frac{Q}{V_m}\left(\frac{\lambda}{2n}\right)^3$, where $V_m$ is the mode volume ($V_m = V$), we get $\rho_{min} = \frac{\sqrt{M}}{1+\sqrt{M}}\frac{\omega\tau_0}{Q^2}\frac{\pi^2}{6}\left(\frac{\lambda}{2n}\right)^{-3}$.

## 1.2 The effects of spatial and spectral inhomogeneity

So far, we have assumed that all emitters interact equally with the optical modes. However, if the gain medium occupies a large volume than the mode, those emitters outside the mode volume will not contribute to stimulated emission but generate spontaneous emission to all $M$ modes. On the other hand, in bulk semiconductors the free electrons and holes can diffuse from high concentration regions to low concentration regions. This carrier diffusion effectively allows those emitters initially outside the mode volume to contribute to stimulated emission into the mode. This effect is neglected.

Likewise, if the emitters have a spectral inhomogeneous broadening width, $\Delta\omega$, that is much larger than the bandwidth, $\Delta\Omega$, of each mode, where $\Delta\Omega = \omega/Q$, not all the emitters contribute to the stimulated emission of the lasing mode although they are located within the mode volume. This spectral overlap effect increases $\rho_{min}$ by a factor of $\frac{\Delta\omega}{\Delta\Omega}$. Taken this into account, we obtain an expression. $\rho_{min} = \frac{\Delta\omega\tau_0}{Q}\frac{\pi^2}{6}\left(\frac{\lambda}{2n}\right)^{-3}\frac{\sqrt{M}}{1+\sqrt{M}}$.

Furthermore, if emitters are randomly oriented with respect to the electric field directions of the modes, an averaging over the polarization (spin) states should be applied. In this case, the required emitter density is increased by a factor of 2. We obtain the following equation:

$$\rho_{min} = \frac{\Delta\omega\tau_0}{Q}\frac{\pi^2}{3}\left(\frac{\lambda}{2n}\right)^{-3}\frac{\sqrt{M}}{1+\sqrt{M}} \qquad (6)$$

## 1.3 The effect of valence-band absorption (transparency condition)

In a bulk semiconductor, electrons in the valence band can induce significant optical absorption of the oscillating modes. This absorption process can be incorporated into the rate equations by adding a term, $-\frac{N_{tr}q}{\tau_s}$, where $N_{tr}$ represents the number of absorbers in the interaction volume $V$, into Eq. (2).

The threshold power becomes:

$$P_{th} = \frac{M}{\tau_c} + \frac{M}{\tau_s}N_{tr} \qquad (7)$$

where $\rho_{tr} = N_{tr}/V$ denotes a transparency emitter density, which is nearly the same the charge carrier density at the transparency condition (see Supplementary Note 3). In III-V semiconductors, $\rho_{tr} \approx 2 \times 10^{18}$ for modes near the band edge and can be $> 5 \times 10^{18}$ when the mode is positioned significantly farther away from the band edge. Therefore, $P_{th} \approx \frac{M\omega_0}{Q} + \frac{MQ}{\tau_o}\rho_{tr}\left(\frac{\lambda}{2n}\right)^3 \frac{6}{\pi^2}$, where $\frac{1}{\tau_s} = \frac{F_p}{\tau_o}$ has been used. The ratio of the second term to the first term is $\frac{Q^2}{\omega_0 \tau_o}\rho_{tr}\left(\frac{\lambda}{2n}\right)^3 \frac{6}{\pi^2}$. For $\rho_{tr} = 5 \times 10^{18}\ cm^{-3}$, $\frac{\lambda}{2n} = 180\ nm$, $\omega_0 = 1.5 \times 10^{15}\ Hz$, and $\tau_o = 10^{-9}\ s$, we find the ratio to be $\frac{Q^2}{85}$. For half-wavelength lasers with typical Q values of 10, both terms contribute approximately equally to the threshold power.

The minimum emitter density is:

$$\rho_{min} \approx \left(\frac{\tau_s}{\tau_c} + \rho_{tr}\right)\frac{q_{th}}{1 + q_{th}} \quad (8)$$

Note that $\rho_{tr}$ is added to the solution of the absorption-less (four-level) case in Eq. (5).

**Supplementary Note 2. Density of states and light amplification in bulk semiconductors**

*2.1 Density of states*

In a bulk 3D semiconductor, the density of states expressed in terms of energy beyond the band edge is $\rho_{e,h}(E)dE = \frac{1}{2\pi^2}\left(\frac{2m_{e,h}^*}{\hbar^2}\right)^{\frac{3}{2}} E^{0.5}\ dE$, where $E = E_{e,h} - E_{c,v}$ is the excess energy of the electron (e) or hole (h) relative to the conduction (c) or valence (v) band edge, and $m_{e,h}^*$ is the effective mass of the carrier. See Extended Fig. 5a for the energy diagram of the states.

The reduced density of states is derived from the carrier density, which reads in terms of the electron-hole pair energy, $E_e + E_h - (E_c + E_v) = \hbar\omega - E_g$, as

$$\rho_E(E) \approx \frac{1}{2\pi^2}\left(\frac{2m^*}{\hbar^2}\right)^{1.5} \sqrt{E} \quad (9)$$

where $E = \hbar\omega - E_g$, $E_g = E_c + E_v$ is the band gap energy, $\hbar\omega$ is the photon (polariton) energy, $\hbar$ is reduced Planck constant, and $m^* = \frac{m_e^* m_h^*}{m_e^* + m_h^*}$ is reduced mass. For In$_{0.53}$ Ga$_{0.47}$As$_{0.92}$P$_{0.08}$, $m_e^* = 0.0584\ m_0$, $m_h^* = 0.505\ m_0$, $m_0 = 9.1 \times 10^{-31}\ kg$, and $E_g = 0.788\ eV$ ($\lambda = 1574\ nm$).

The total number of states in this gain bandwidth is:

$$\rho_{tot} = \int_0^{\Delta\omega} \rho_E(E)dE = \frac{1}{3\pi^2}\left(\frac{2m^*\hbar\Delta\omega}{\hbar^2}\right)^{1.5} \quad (10)$$

Under optical pumping using 1064-nm laser light ($E_p = 1.165\ eV$), the width of the excited emitter states is $\hbar\Delta\omega = 0.38\ eV$. The maximum number of states is $9.7 \times 10^{18}\ cm^{-3}$ from 1574 to 1064 nm. Consider a single cavity mode with a resonance frequency $\omega_0\ (= \frac{2n\pi c}{\lambda})$ and a width of $\Delta\Omega\ (= \frac{\omega_0}{Q})$. The number of states to fill up for the quasi-Fermi level to reach $1240\ nm\ (\lambda)$ is $4.1 \times 10^{18}\ cm^{-3}$. For $\hbar\Delta\Omega = 0.1$ eV ($Q = 10$), the number of states within the modal resonance bandwidth is $2.9 \times 10^{18}\ cm^{-3}$.

### 2.2 Semiconductor gain constant

Consider a plane wave with a frequency $\omega$ and intensity $I_\omega$ propagating through a gain medium with $\rho_\omega(\omega)$ elements per unit volume per frequency. The intensity change over distance is described as $I(z) = I(0)e^{G(\omega)z}$ using a gain constant $G = \frac{\pi c^2}{2n^2\omega^2\tau_s}\rho_\omega(\omega)$ where $\tau_s$ denotes the spontaneous relaxation time of the emitters. In bulk semiconductors, the emitters have two polarization states. For free-space amplification, only a half of the emitters with the polarization state same as that of the wave contribute to amplification. In this case, the single polarization gain $G_{pol}$ should be a half of $G$. Using the relation $\rho_\omega(\omega) = \rho_E(E)\hbar$, we get

$$G_{pol} = \left(\frac{\lambda}{2n}\right)^2 \frac{\hbar}{4\pi\tau_s} \rho_E(E) \quad (11)$$

For the example above and assuming $\tau_s = 10^{-9}\ s$ and $n = 3.5$, we obtain $G_{pol} = 2.69 \times 10^3\ cm^{-1}$.

In a laser cavity, threshold is reached when unsaturated gain equals loss: that is, $G = \frac{n}{c\tau_c} = \frac{2n\pi}{\lambda Q}$. For $Q = 10$, $G = 1.75 \times 10^4\ cm^{-1}$. This gain value required for threshold is about 6.5 times greater than the free-space gain of the semiconductor. Therefore, it is essential to take advantage of reduced spontaneous time $\tau_s$ via the Purcell enhancement: that is, $F_p > 6.5$.

### 2.3 Nonradiative recombination rates in semiconductor

The spontaneous emission rate in a semiconductor with $n\ (= p)$ numbers of electrons (holes) per unit volume is proportional to the carrier density: $\frac{1}{\tau_s} = Bn$, $B$ is a recombination coefficient. As $n$ increases, the non-radiative Auger recombination rate increases: $\frac{1}{\tau_{nr}} = Cn^2$, where $C$ is the Auger recombination coefficient. The total decay is

$$\frac{1}{\tau} = \frac{1}{\tau_s} + \frac{1}{\tau_{nr}} = Bn + Cn^2 \quad (12)$$

For undoped InGaAsP at 1.3 μm and room temperature, the typical coefficient values are $B = 2 \times 10^{-10}\ cm^3/s$ and $C = 2.3 \times 10^{-29}\ cm^6/s$. For $n = 5 \times 10^{18}\ cm^{-3}$, we obtain $\tau_s = 1\ ns$ and $\tau_{ns} = 1.7\ ns$, and for $\rho = 10^{19}\ cm^{-3}$, $\tau_s = 0.5\ ns$ and $\tau_{ns} = 0.43\ ns$. Extended Figure 5a shows a plot of $\frac{Bn}{Bn+Cn^2}n$ as a function of $n$ for four different $B$ values and $C = 2.3 \times 10^{-29}\ cm^6/s$. For $B = 2 \times 10^{-10}\ cm^3/s$, it is not possible to fill the excited states more than $6 \times 10^{18}\ cm^{-3}$.

However, in a plasmonic nano-cavity the radiative coefficient emitter $B$ can be significantly enhanced by the Purcell effect. For example, for $B = 5 \times 10^{-19}\ cm^3/s$, the Auger recombination loss is nearly negligible at $n = 6 \times 10^{18}\ cm^{-3}$ and modest even at $n = 1.3 \times 10^{19}\ cm^{-3}$, the maximum number of emitters achievable with 1064 nm pumping.

**Supplementary Note 3. The semiconductor laser rate equation model in the spectral domain**

We wish to write rate equations in the spectral domain in terms of the spectral densities of emitters, $n_i(t)$, and of photons, $q_j(t)$. $n_i$ and $q_i$ represent the numbers of emitters and photons in the cavity within a frequency band from $\omega_i$ to $\omega_{i+1}$, where $\omega_{i+1} - \omega_i = \delta\omega$. Let us first consider a mode with a square spectral profile with a center at $\omega_0$ and a width of $\Delta\Omega$ ($\delta\omega \ll \Delta\Omega$). The total number of photons within the optical bandwidth is $q(t) = \sum q_j(t)$, and $q_j(t) = \frac{\delta\omega}{\Delta\Omega}q(t)$ for $\omega_i \in [\omega_0 - \frac{\Delta\Omega}{2}, \omega_0 + \frac{\Delta\Omega}{2}]$. The total number of emitters within this bandwidth is $N_1 = \sum n_i(t)$, and $n_j(t) = \frac{\delta\omega}{\Delta\Omega}N_1(t)$. Let us assume that the emitter bandwidth is wider than the modal bandwidth so that $n_j(t)$ is nearly constant for $\omega_j \in [\omega_0 - \frac{\Delta\Omega}{2}, \omega_0 + \frac{\Delta\Omega}{2}]$. The interaction of single-mode photons and emitters can be described by (see Supplementary Note 1):

$$\frac{dN_1(t)}{dt} = P(t) - \frac{N_1 q}{\tau_s} - \frac{N_1}{\tau_s}$$

$$\frac{dq(t)}{dt} = \frac{N_1 q}{\tau_s} + \frac{N_1}{\tau_s} - \frac{q}{\tau_c}$$

Inserting $N_1 = \sum n_i(t)$ and $q(t) = \sum q_j(t)$ and using $\sum q_j(t) = \frac{\Delta\Omega}{\delta\omega}q_i(t)$ and $\sum n_j(t) = \frac{\Delta\Omega}{\delta\omega}n_i(t)$, we obtain:

$$\frac{d\sum n_i(t)}{dt} = P(t) - \frac{\Delta\Omega}{\delta\omega}\frac{\sum n_i(t)\,q_i(t)}{\tau_s} - \frac{\sum n_i(t)}{\tau_s}$$

$$\frac{d\sum q_j(t)}{dt} = \frac{\Delta\Omega}{\delta\omega}\frac{\sum n_i(t)\,q_i(t)}{\tau_s} + \frac{\sum n_i(t)}{\tau_s} - \frac{\sum q_j(t)}{\tau_c}$$

For a realistic Lorentzian profile mode, where $q_i(t) = q(t)\frac{\delta\omega}{\pi}\frac{\Delta\Omega/2}{(\omega_i-\omega_o)^2+\Delta\Omega^2/4}$, the cross-product term cannot be simply decomposed as above. Using physical insights based on the level diagram depicted in Extended Figure 3d, we may write spectral-domain rate equations as in the following.

For emitters in levels labeled from 1 (lowest) to $k$ (highest, pump level),

$$\frac{dn_k}{dt} = P_k(t) - \frac{n_k}{\tau_s(\omega_k)} - \frac{n_k}{\tau_{nr}(\omega_k)}$$

$$\frac{dn_{k-1}}{dt} = -B\frac{n_{k-1}q_{k-1}}{\tau_s(\omega_{k-1})} - \frac{n_{k-1}}{\tau_s(\omega_{k-1})} - \frac{n_{k-1}}{\tau_{nr}(\omega_{k-1})}$$

$$\frac{dn_{k-2}}{dt} = -B\frac{n_{k-2}q_{k-2}}{\tau_s(\omega_{k-2})} - \frac{n_{k-2}}{\tau_s(\omega_{k-2})} - \frac{n_{k-2}}{\tau_{nr}(\omega_{k-2})}$$

$$\ldots$$

$$\frac{dn_1}{dt} = -B\frac{n_1 q_1}{\tau_s(\omega_1)} - \frac{n_1}{\tau_s(\omega_1)}$$

For photons (polaritons) and $i = [1, k-1]$:

$$\frac{dq_i}{dt} = B\frac{n_i q_i}{\tau_s(\omega_i)} + \frac{n_i}{\tau_s(\omega_i)} - \frac{q_i}{\tau_c(\omega_i)}$$

Thermal excitation was neglected to reduce computational time.

The rate constant parameters are given as a function of frequency as follows:

$$B = \frac{\Delta\Omega}{\delta\omega}$$

$$\frac{1}{\tau_s(\omega_i)} = \frac{1}{\tau_s} \cdot \frac{\Delta\Omega^2/4}{(\omega_i - \omega_o)^2 + \Delta\Omega^2/4}$$

$$\frac{1}{\tau_c(\omega_i)} = \frac{1}{\tau_{c0}} / \frac{\Delta\Omega^2/4}{(\omega_i - \omega_o)^2 + \Delta\Omega^2/4}$$

$$\frac{1}{\tau_{nr}(\omega_i)} = \frac{1}{\tau_{nr0}} \cdot \frac{n_{sat}(\omega_{i-1}) - n_{i-1}}{n_0}$$

The coefficient $B$ plays a central role in these spectral-domain rate equations. It makes the simulation result independent of the choice of $\delta\omega$. The $\Delta\Omega$ as the numerator has been derived from the basic 3- or 4-level rate equations, which represents the cold-cavity modal bandwidth. A question may be asked whether the numerator should be the actual spectral linewidth of the mode, which varies as the spectrum evolves through lasing threshold. We used a constant value of $\Delta\Omega$ ($= \omega/Q$) in the computer simulation.

To avoid unrealistically large $\tau_s(\omega)$, we limit the maximum $\tau_s$ to be $\tau_0$. To avoid an excessive computational time, a time interval of 3.3 fs was typically used. This value was adequate for simulating $Q = 10$ modes, for which $\tau_{c0} = 6.6\ fs$, and the minimum value of $\tau_c(\omega_i)$ was limited to the time interval value. $\tau_{nr}(\omega_i)$ is a time constant for nonradiative level-filling decays from the higher to lower excited levels. The decay speed is proportional to the number of empty states in the lower levels, represented by $\frac{n_{sat}-n_{i-1}}{n_0}$, where $n_{sat}$ is the maximum number of states in the $i$-th level:

$$n_{sat}(\omega_i) = n_0 \left(\frac{\omega_i - \omega_1}{\Delta\omega}\right)^{0.5}$$

where $n_0 = \frac{3}{2}\frac{\delta\omega}{\Delta\omega}\frac{1}{6\pi^2}\left(\frac{2m^*}{\hbar^2}\right)^{1.5}(\hbar\Delta\omega)^{1.5}$. Note that $\sum_1^k n_{sat}(\omega_i) = 0.5\,\rho_{tot}$; that is, we assumed optical gain to come from only a half of the total number in Eq. (10) due to the polarization matching (spin conservation) rule. However, the gain elements are expected to undergo spatial diffusion and polarization randomization in the semiconductor gain medium. Therefore, the effective number of gain elements contributing to gain may actually be greater than $0.5\,\rho_{tot}$.

The pump rate has a Gaussian pulse shape given by

$$P_k(t) = N_p \rho_{tot} 2/\sqrt{\pi}\, e^{-4t^2/\tau_p^2}$$

where $N_p$ corresponds to the total number of absorbed pump photons normalized to the number of states, and $\tau_p$ is the 1/e full width of the pump pulse.

Once all the parameters are assigned with reasonable values, we found that the simulation results depend on the coefficient of the stimulated emission term, $\frac{\hbar\Delta\Omega}{\tau_s}$. We used this coefficient as the main fitting parameter to reproduce the experimental laser mode spectra as closely as possible. For half-wavelength dipole modes, best fitting results were obtained using $\tau_s = 10 - 11\,ps$ at the mode center frequency. The relative magnitude of fluorescence background with respect to the lasing peak could be adjusted by using $\tau_0$ as a free parameter. In the simulation, $\tau_0 = 0.2\,ns$ was used. This corresponds to $F_p = \frac{\tau_0}{\tau_s} = 18 - 19$ (see below).

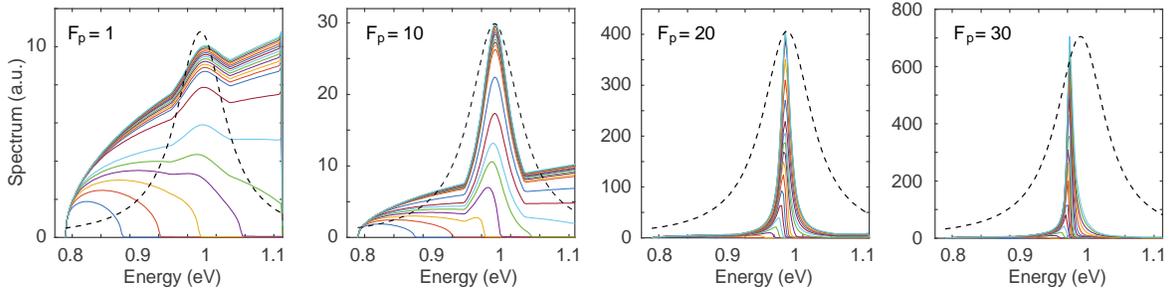

**Supplementary Note 4. The resonance and coupling of the plasmonic and dielectric modes**

    (i)    A heuristic comparison of the electrical dipole (ED) modes

Analytic solutions exist for Mie scattering from spheres. The ED mode resonance of a metallic sphere (perfect conductor or noble metals away from the surface plasmon frequency) is:

$$\lambda_{ED} \approx 5.62\, n_m R$$

where $R$ is the radius and $n_m$ is the refractive index of the surrounding medium.

On the other hand, a dielectric sphere with a refractive index $n_1$ ($> n_m$) has the ED mode resonance at

$$\lambda_{ED} = 1.25\, (n_1 + n_m)R.$$

We find that $\lambda_{ED}$ (gold) = $\lambda_{ED}$ (dielectric) is met for the same $R$ when $n_1 = 3.50\, n_m$. For instance, a semiconductor sphere of $n_1 = 3.5$ has identical ED resonance wavelength as a perfect conductor sphere in the air ($n_m = 1$).

Since disks and spheres have similar ED resonance wavelengths, it is well expected that the ED resonance wavelengths of a gold disk and an InGaAsP semiconductor disk are closely matched.

    (ii)    Resonance modes of various metal and semiconductor nano-disks and their hybrid structures

Figures S2 below display the Mie scattering spectra of various device sizes, obtained by FDTD simulation. The ED plasmonic mode emerges with a gold thickness as thin as 10 nm, less than half the skin depth.

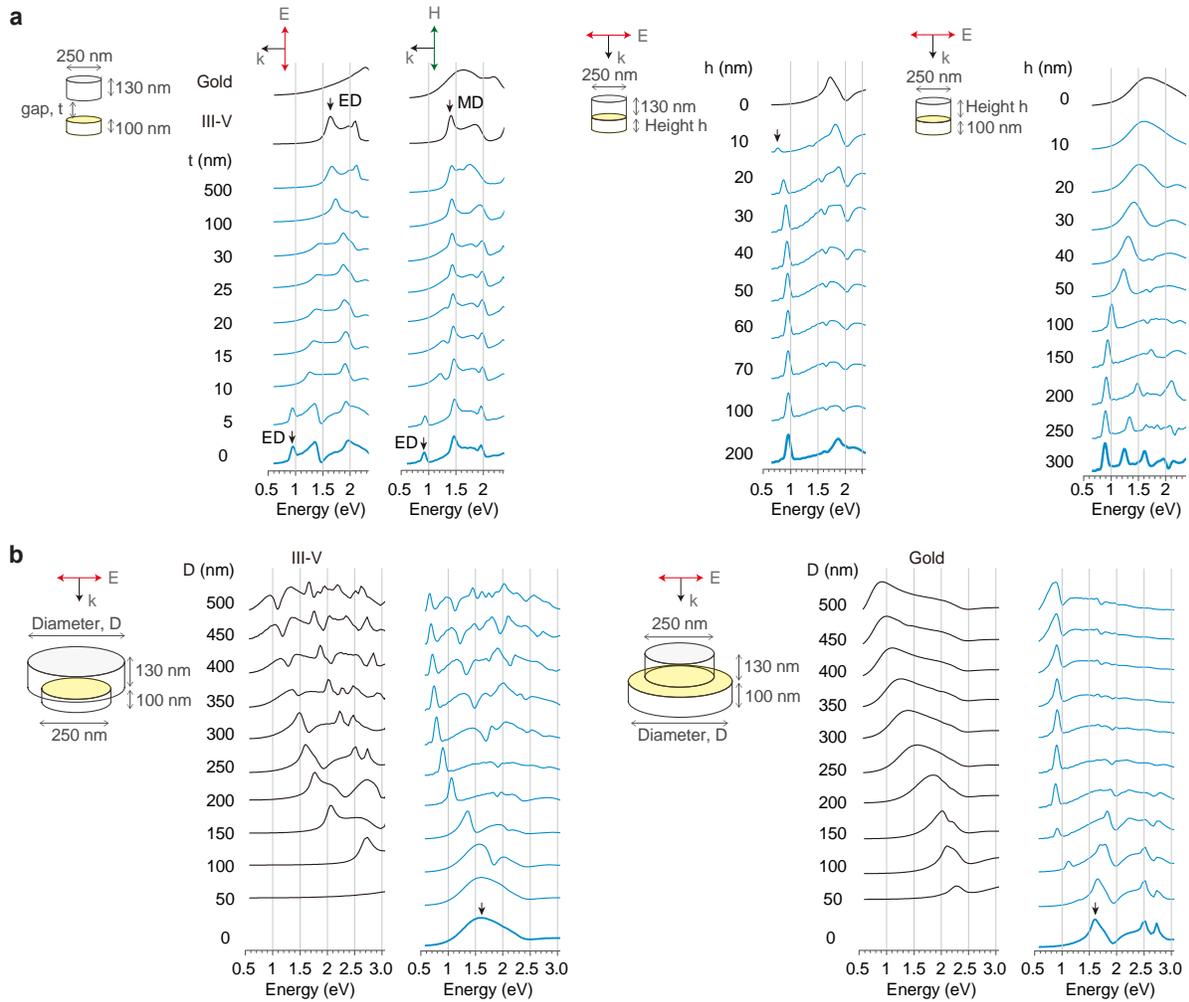

**Figure S2. Hybridization of semiconductor and plasmonic disks** at **different sizes. a,** Simulated Mie scattering spectra of the hybridization between semiconductor (Diameter D = 250 nm, height h = 130 nm) and plasmonic disks (D = 250 nm, h = 100 nm) at different gap distances (t = 0~500 nm) are shown, with illumination from the side under two different polarizations (Left: Transverse Magnetic - TM, Right: Transverse Electric - TE). **b,** Simulated Mie scattering spectra of the hybridization between semiconductor and plasmonic disks at different sizes are depicted. On the left, the semiconductor has varying diameters (D = 0~500 nm) with a constant height (H = 130 nm), while the right side shows fixed gold disk size (D = 250 nm, h = 100 nm) with varying semiconductor diameter. The hybridization occurs with zero gap distance (t = 0) and is illuminated from the top with Transverse Electric (TE) polarization.

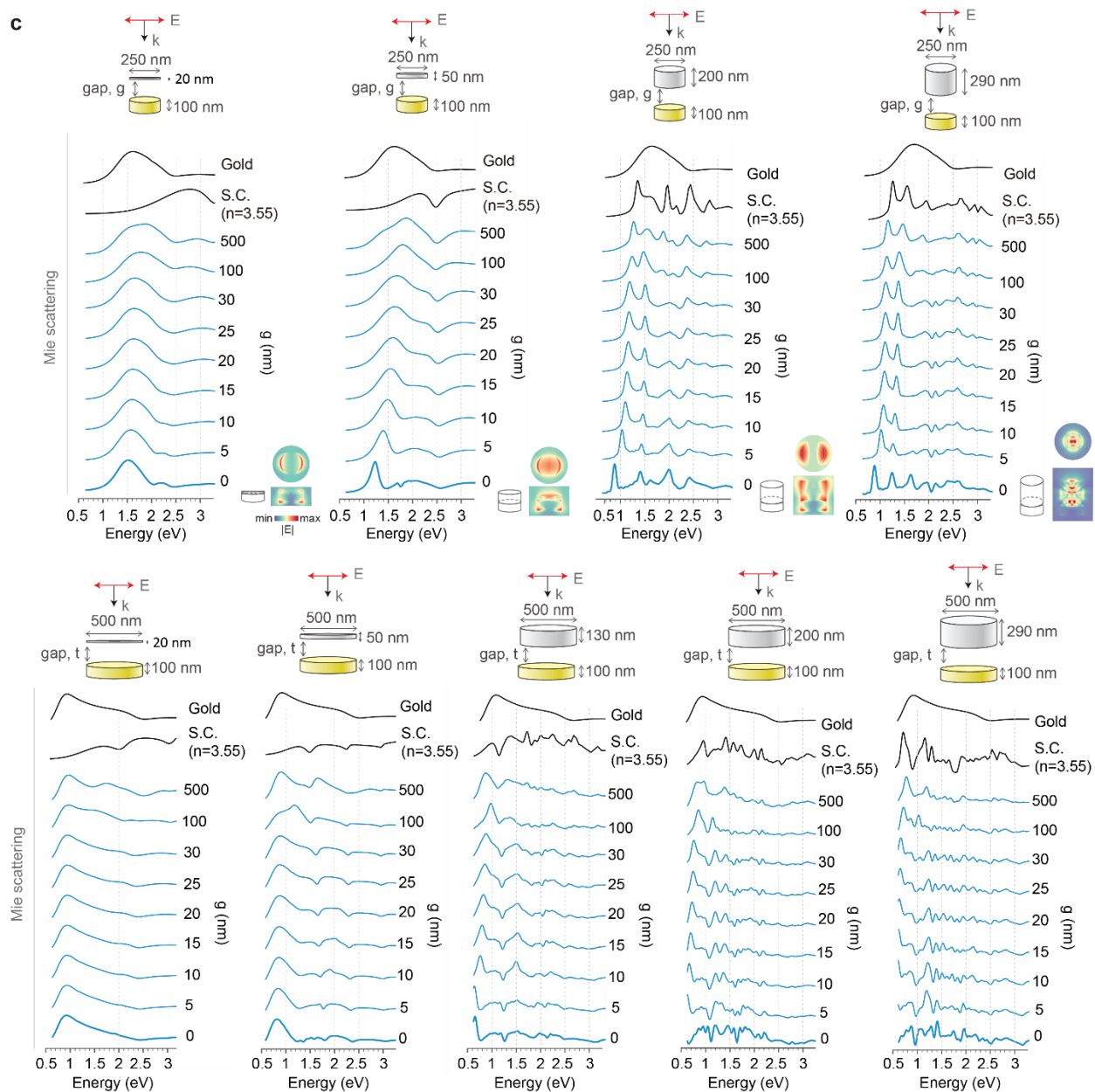

**Figure S2 (continued). c,** Simulated Mie scattering spectra of the hybridization between semiconductor and plasmonic disks at different sizes are depicted. On the left, the semiconductor has varying diameters (D = 0 to 500 nm) with a constant height (h = 130 nm), while the right side shows fixed gold disk size (D = 250 nm, h = 100 nm) with varying semiconductor diameter. The hybridization occurs with zero gap distance (g = 0) and is illuminated from the top with Transverse Electric (TE) polarization.

**Supplementary Note 5. Finite difference time domain (FDTD) simulation**

FDTD simulations were performed using commercial software (Lumerical). We used the refractive index data of the gold substrate we have previously measured by using optical ellipsometry. The constant real part of refractive index of InGaAsP semiconductors was assumed to be constant at 3.55.

(i) Mie scattering simulation

We employed a total-field scattered-field (TFSF) source to illuminate samples with a plane wave. The scattered field is separated from the incident field, from which scattering spectra. To measure the scattered net power, a power monitor encompassing the sample was used. The polarization and direction of incident wave was varied to calculate TE and TM mode resonances.

(ii) Dipole simulation

An electric or magnetic dipole was placed inside semiconductor particles to excite resonance modes. The lowest order plasmonic mode was efficiently excited by an electric dipole oriented parallel to the metal substrate. To measure the time-dependent electric and magnetic fields, point-like time monitors positioned densely inside or near the semiconductor particles were used.

A fast Fourier transform (FFT) of the acquired data produced spectral responses of the dipoles. From the resonance frequencies ($\varpi_{res}$) and the full width half maximum (FWHM) ($\Delta\varpi_{res}$) spectral widths, the Q factors of low-Q modes were calculated: $Q = \varpi_{res}/\Delta\varpi_{res}$. For high Q modes that do not decay completely within the simulation timeframe, their Q values were determined from the slope of their electric field decay profiles.

The three-dimensional near-field field pattern of a mode was obtained using an array of two-dimensional field monitors over the entire simulation region. The polarization charge of a mode was calculated by taking the divergence of the electric field recorded in the monitors. A proper apodization was applied to remove the transient effect of the dipole source. The far-field emission pattern of a device was computed using a box monitor that encompassed the entire metal-semiconductor structure.

(iii) Mesh settings

For mesh refinement, the conformal variant 2 setting was used utilizing the Yu-Mittra method for interfaces involving metals. We employed the perfectly matched layer (PML) as the boundary condition. The simulation volume was set to be three times larger than a particle of interest. The mesh size was kept below 5 nm for sufficient accuracy.

**Supplementary Data Table 1. Representative room-temperature nano-lasers to date.**

| | Cavity structure | Gain material (Wavelength) | Particle size $(\lambda/2n)^3$ | Mode type | Fab method | Threshold energy (Cal.) | Pulse duration (Rep.) | Peak power (Cal.) | Year (Ref) |
|---|---|---|---|---|---|---|---|---|---|
| R1 | Disk (on pillar) | InGaP quantum well, InGaAlP clad (645 nm) | 645 nm (dia) 180 nm (height) 73 $(\lambda/2n)^3$ | WGM | E-beam lithography | (70 mJ/cm$^2$) | 8 ns (33 kHz) | 50 µW, optical (8.8 MW/cm$^2$) | 2007 ([1]) |
| R2 | Rect pillar (silver-coated, on substrate) | InGaAs (1500 nm) | 310 nm (length) 6 µm (length) 300 nm (height) 59 $(\lambda/2n)^3$ | Fabry-Perot | E-beam lithography | 5.9 mA, electrical | 28 ns (1 MHz) | | 2009 ([2]) |
| R3 | Hexagonal nanodisk (dielectric) | ZnO on silica (390 nm) | 612 nm (dia) 550 nm (height) 286 $(\lambda/2n)^3$ | WGM | Chemical vapor transport growth | (2 mJ/cm$^2$) | 5 ns (10 Hz) | 0.4 MW/cm$^2$, optical | 2010 ([3]) |
| R4* | Disk (aluminum-coated, on substrate) | InGaAsP quantum well (1430 nm) | 1100 nm (dia) 1350 nm (height) 31 $(\lambda/2n)^3$ | WGM | E-beam lithography | (840 µJ/cm$^2$) | 12 ns (300 kHz) | 700 W/mm$^2$, optical | 2010 ([4]) |
| R5 | Hexagonal nanopillar (dielectric) | InGaAs core in GaAs shell (950 nm) | 540 nm (dia) 3 µm (height) 267 $(\lambda/2n)^3$ | Helical | Nanowire growth | 93 µJ/cm$^2$, optical | 120 fs (76 MHz) | (775 MW/cm$^2$) | 2011 ([5]) |
| R6 | Square plates (on silver substrate) | CdS (500 nm) | 1000 nm (side) 45 nm (thick) 45 $(\lambda/2n)^3$ | WGM | Vapor-phase growth | (200 µJ/cm$^2$) | 100 fs (10 kHz) | 2000 MW/cm$^2$, optical | 2011 ([6]) |
| R7 | Nanowires (on aluminum substrate) | GaN (375 nm) | 100 nm (side) 15 µm (length) 28 $(\lambda/2n)^3$ | Fabry-Perot | Vapor-phase growth | (35 mJ/cm$^2$) | 10 ns (100 kHz) | 3.5 MW/cm$^2$, optical | 2014 ([7]) |
| R8 | Square Plates (on gold substrate) | CdSe (700 nm) | 1000 nm (side) 137 nm (height) 50 $(\lambda/2n)^3$ | WGM | Vapor-phase growth | (450 µJ/cm$^2$) | 4.5 ns (1 kHz) | 100 kW/cm$^2$, optical | 2017 ([8]) |
| R9 | Disk (free-standing) | InGaAsP (1170~1580 nm) | 1800 nm (dia) 250 nm (height) 100 $(\lambda/2n)^3$ | WGM | Optical lithography | 7 pJ, optical (222 µJ/cm$^2$) | 3 ns (2 MHz) | (74 kW/cm$^2$) | 2019 ([9]) |
| R10 | Disk (free-standing) | InGaP/InAlGaP quantum well (670~690 nm) | 700 nm (dia) 180 nm (height) 86 $(\lambda/2n)^3$ | WGM | E-beam lithography | 30 µJ/cm$^2$, optical | 1.5 ns (100 Hz) | (20 kW/cm$^2$) | 2019 ([10]) |
| R11 | Cube (on dielectric substrate) | CsPbBr$_3$ (530 nm) | 310 nm (side) 19 $(\lambda/2n)^3$ | Mie | Solution deposition | 300 µJ/cm$^2$, optical | 150 fs (100 kHz) | (2000 MW/cm$^2$) | 2020 ([11]) |
| R12 | Disks (on sodium substrate) | InGaAsP multi quantum well (1257 nm) | 1200 nm (dia) 200 nm (height) 41 $(\lambda/2n)^3$ | WGM | Electron-beam lithography | (700 µJ/cm$^2$, optical) | 5 ns (12 kHz) | 140 kW/cm$^2$ | 2020 ([12]) |
| R13 | Cubes (on gold substrate) | CsPbBr$_3$ (540 nm) | 580 nm (side) 320 nm (height) 67 $(\lambda/2n)^3$ | WGM | Sono-chemistry | 700 µJ/cm$^2$, optical | 5 ns (20 Hz) | (140 kW/cm$^2$) | 2021 ([13]) |
| R14 | Square plates (on silver substrate) | CsPbBr$_3$ (540 nm) | 1060 x 830 nm$^2$ 77 nm (height) 42 $(\lambda/2n)^3$ | WGM | Solution deposition | 26 µJ/cm$^2$, optical | 190 fs (80 MHz) | (137 MW/cm$^2$) | 2022 ([14]) |
| R15 | Disk (on pillar) | InGaP bulk (650 nm) | 460 nm (dia) 200 nm (height) 43 $(\lambda/2n)^3$ | WGM | Optical lithography | 1 mJ/cm$^2$, optical | 5 ns (20 Hz) | (0.2 MW/cm$^2$) | 2023 ([15]) |
| R16 | Disk (on gold substrate) | InGaP bulk (650 nm) | 360 nm (dia) 200 nm (height) 27 $(\lambda/2n)^3$ | WGM | Optical lithography | 0.9~1.5 mJ/cm$^2$, optical | 5 ns (20 Hz) | (0.18~0.3 MW/cm$^2$) | 2023 ([15]) |
| R17 | Rhombus (on gold substrate) | InGaAsP bulk (1190~1460 nm) | 190 nm (side) 130 nm (height) 0.7 $(\lambda/2n)^3$ | Fabry-Perot | Optical lithography & Wet etching | 0.2~0.9 mJ/cm$^2$, optical | 2ns (2.5 MHz) | (0.1~0.45 MW/cm$^2$) | This work |

*R4: The gain medium core has a diameter of 980 nm (in the paper, the radius of the core was referred as major diameter) and an estimated height of 480 nm, which is computed from 1300 nm (total) minus 500 nm (core to the top

side of SiO$_2$) minus 250 nm (core to the glass window), and minus 70 nm (metal shield)). The effective index (n) was estimated to be 1.84 by volume averaging, taking into account the SiO$_2$ cap (with $n_{sio2}$=1.46, a thickness of 200 nm, and a total height of 1230 nm, including 500 nm for the InGaAsP core, 250 nm for the SiO$_2$ layer, and 480 nm for the gain core) and the InGaAsP core ($n_{InGaAsP}$=3.4).

**Supplementary Data Table 2. Representative cryogenic-temperature nano-lasers to date.**

|  | Cavity structure | Gain materials (wavelength) | Particle Size $(\lambda/2n)^3$ | Mode type | Fab Method | Temperature | Threshold (cal.) | Pulse duration (rep.) | Peak power (cal.) | Year (ref) |
|---|---|---|---|---|---|---|---|---|---|---|
| C1* | Rect. pillar (gold-coated, on substrate) | InGaAsP (1400 nm) | 260 nm (dia), 500 nm (height) 3.3 $(\lambda/2n)^3$ | Fabry-Perot | E-beam lithography | 77 K | 4 µA, electrical | CW |  | 2007 ([16]) |
| C2 | Disk (on pillar) | InAs QD in GaAs (860 nm) | 627 nm (dia), 265 nm (height) 46 $(\lambda/2n)^3$ | WGM | Optical lithography, wet etching | 10 K | (50 µW, optical) | 200 fs (76 MHz) | 0.02 MW/cm$^2$ | 2009 ([17]) |
| C3 | Nanowires (on silver substrate) | CdS (489 nm) | 129 nm (dia) 12 µm (length) 24 $(\lambda/2n)^3$ | Fabry-Perot | Vapor-phase growth | < 10K | 500 nJ/cm$^2$ | 100 fs (80 MHz) | 5 MW/cm$^2$, optical | 2009 ([18]) |
| C4 | Disk (silver-coated, on substrate) | InAsP QWs (1420 nm) | 1000 nm (dia) 235 nm (thick) 23 $(\lambda/2n)^3$ | WGM | E-beam lithography | 8K | (120 kW/cm$^2$) | 60 ps (80 MHz) | 120 kW/cm$^2$ | 2010 ([19]) |
| C5* | Coaxial (silver-coated, on substrate) | InGaAsP quantum well (1427nm) | 200 nm (core), 100 nm (ring thicknes) 200 nm (height) 1.96 $(\lambda/2n)^3$ | WGM | E-beam lithography | 4.5 K | - | CW | - | 2012 ([20]) |
| C6* | Nanowires (on silver substrate) | InGaN@GaN core-shell (510 nm) | 650 nm (length), 30 nm (hexa-side) 1.3 $(\lambda/2n)^3$ | Fabry-Perot | Molecular beam epitaxy | 78K | (3.7 kW/cm$^2$) | CW | 3.7 kW/cm$^2$, optical | 2012 ([21]) |
| C7* | Nanowires (on silver substrate) | InGaN@GaN core-shell (554 nm) | 240 nm (length), 42 nm (dia) 0.37 $(\lambda/2n)^3$ | Fabry-Perot | Molecular beam epitaxy | 7 K |  | CW | 100 W/cm$^2$, optical | 2014 ([22]) |
| C8 | Nanocylinder | GaAs on quartz (825 nm) | 500 nm (dia) 330 nm (height) 41 $(\lambda/2n)^3$ | Quasi bound states | E-beam lithography | 77 K | (433 µJ/cm$^2$) | 200 fs (20 kHz) | 2.2 MW/cm$^2$, optical | 2020 ([24]) |

*Footnotes

C1: The size of this device is referred to only the gain medium, including a thickness of 25 nm for silicon nitride and an 80 nm-thick gold coating. The device is expected to exhibit an HE11-like mode.

C5: This device used a transverse electromagnetic (TEM)-like mode.

C6: This device consists of an InGaN gain part (170 nm in length) and a GaN part (480 nm in length). Given the refractive index of GaN (~2.45) and the total length of the nanorod (650 nm), the longitudinal mode order is calculated to be 6.

C7: Given the refractive index of GaN (~2.45) and the total length of the nanorod (240 nm), the longitudinal mode order is calculated to be 2. However, the longitudinal mode order was not discussed in this paper.

**Supplementary Data Table 3. Controversial spaser devices**

|  | Cavity structure | Gain materials (wavelength) | Particle Size | Mode type | Fab Method | Temperature | Threshold (cal.) | Pulse duration (rep.) | Peak power (cal.) | Year (ref) |
|---|---|---|---|---|---|---|---|---|---|---|
| D1* | Gold nanoparticles on substrate | Oregon Green 488 (531 nm) | 14 nm (core) 44 nm (silica shell) | LSP Dipole | Wet chemistry | 298K | 0.005 mJ (11 kJ/cm$^2$) | 5 ns (not reported) | 2.2 GW/cm$^2$ | 2009 ([25]) |
| D2* | Gold nanoparticles in water | Organic Uranine dye (550 nm) | 10 nm (core) 6 nm (silica shell) | Not discussed | Wet chemistry | 298 | 26 mJ/cm$^2$ | 7 ns (100 Hz) | (3.7 MW/cm$^2$) | 2017 ([26]) |
| D3* | Silver nanocubes on gold substrate | A single CsPbBr$_3$ quantum dot | 100 nm (side) 100 nm (height) | Gap plasmon | Bottom-up | 120K | (1.9 W/cm$^2$) | CW | (1.9 W/cm$^2$) | 2020 ([23]) |

*Footnotes

D1: This paper has been cited as the first experimental spaser. However, extensive efforts following the paper in the research community, scientists, and the authors themselves, failed to obtain evidence for single particle lasing. Possible reasons why the design is unable to generate single nanoparticle spasing have been proposed.[27] It was suspected that the observed spectral narrowing might be due to random feedback lasing by many nanoparticles.[28] Although single particle spasing is now considered a goal yet to be attained, unfortunately the earlier claims to the contrary have persisted, contributing to misleading statements within the literature. It will be important to rectify the confusion and misconception surrounding spasers.

D2: This paper used dye-coated nanoparticles, quite similar to D1, and claimed to be the first experimental demonstration of spasers in a biological system. However, same concerns to those for D1 are applied.[27] There is a lack of direct experimental evidence for single particle lasing, as all the results were apparently obtained from a colloidal or suspension of nanoparticles. Alternatively, a possible mechanism of whispering gallery mode lasing supported by ultrasonic bubble created by heating of plasmonic nanoparticles has been proposed.[29] Again, it will be important to resolve the controversies.

D3: This paper described the observation of spectral narrowing at cryogenic temperature of a single CsPbBr$_3$ quantum dot emission (10-nm size) that was placed in a narrow gap between a 100 nm-sized silver cube and a gold substrate. Considering the low Q factor of ~ 9 expected for such a configuration,[30,31] the minimum gain constant required for lasing is 13,000 cm$^{-1}$. This level of amplification cannot be obtained from a single quantum dot and even unlikely from a high concentration of quantum dots.[32] Therefore, the observed narrowband emission at low temperature is unlikely due to lasing.

References in Supplementary Tables 1 to 3